\documentclass[useAMS,usenatbib,aas_macros]{mn2e}

\usepackage{epsfig}
\usepackage{lscape,graphicx}
\usepackage{amssymb}
\usepackage{natbib}
\usepackage{times}
\usepackage{aas_macros}

\newcommand{\ifm}[1]{\relax\ifmmode#1\else$\mathsurround=0pt#1$\fi}
\newcommand{\kms}{\ifmmode\,{\rm km}\,{\rm s}^{-1}\else km$\,$s$^{-1}$\fi}

\newcommand{\hmsun}{\,\ifm{h^{-1}}{M_{\odot}}}
\def\omm{\Omega_{\rm m}}
\def\oml{\Omega_{\Lambda}}

\newcommand{\dd}{{\rm d}}

\newcommand{\dS}{\Delta S}
\newcommand{\dW}{\Delta \omega}

\newcommand{\be}{\begin{equation}}
\newcommand{\ee}{\end{equation}}
\newcommand{\bea}{\begin{eqnarray}}
\newcommand{\eea}{\end{eqnarray}}
\newcommand{\z}{\emph{z}}
\newcommand{\nbs}{$N$-body simulation}
\newcommand{\nbss}{$N$-body simulations}

\usepackage{color}

\begin{document}

\title[Universal Merger Histories]
      {Universal Merger Histories of Dark-Matter Haloes}

\author[E.~Neistein, A.~V.~Macci\`{o} \& A.~Dekel]
{Eyal Neistein$^{1,2}$\thanks{E-mail:$\;$eyal@mpa-garching.mpg.de} ,
        Andrea V. Macci\`{o}$^3$ \& Avishai Dekel$^1$
        \\
      $^1$ Racah Institute of Physics, The Hebrew University,
          Jerusalem, Israel \\
$^2$ Max-Planck-Institut f\"{u}r Astrophysik, Karl-Schwarzschild-Str. 1, 85748 Garching, Germany\\
          $^3$ Max-Planck-Institut f\"{u}r Astronomie, K\"{o}nigstuhl 17, 69117 Heidelberg, Germany }


\date{}
\pagerange{\pageref{firstpage}--\pageref{lastpage}} \pubyear{2009}
\maketitle

\label{firstpage}


\begin{abstract}
We study merger histories of dark-matter haloes in a suite of $N$-body
simulations that span different cosmological models. The simulated cases
include the up-to-date WMAP5 cosmology and other test cases based on the
Einstein-deSitter cosmology with different power spectra. We provide a robust
fitting function for the conditional mass function (CMF) of progenitor haloes
of a given halo.  This fit is valid for the different cosmological models and
for different halo masses and redshifts, and it is a significant improvement
over earlier estimates.
Based on this fit, we develop a simple and accurate
technique for transforming the merger history of a given simulated halo into
haloes of different mass, redshift and cosmology. Other statistics such as
main-progenitor history and merger rates are accurately transformed as well.
This method can serve as a useful tool for studying
galaxy formation.
It is less sensitive to the low accuracy of the fit at small time-steps, and it
can thus replace the more elaborate task of construction Monte-Carlo
realizations.
As an alternative approach, we confirm the earlier finding by
Neistein \& Dekel that the main-progenitor follows a log-normal
distribution.  This property of merger trees allows us to better capture their
behaviour as a function of time and descendant mass, but a
broader suite of simulations is required for evaluating the dependence of the
log-normal parameters on the cosmological model.
\end{abstract}


\begin{keywords}
cosmology: theory --- dark matter --- galaxies: haloes --- galaxies:
formation --- gravitation
\end{keywords}


\section{Introduction}
\label{sec:intro}

The growth of dark-matter haloes through merging and accretion is
driving the formation and evolution of galaxies.
An accurate theoretical prediction for the way haloes grow is thus a
key element in the effort to develop a theoretical understanding
of the processes associated with galaxy formation,
including star formation, the growth of black holes in galaxy centers
and their appearance as quasars.

The conditional mass function (hereafter CMF) has been an important
tool in quantifying the growth of haloes. It is defined at a given time
as the average number of progenitors that will merge into a descendant
halo at a later time. The CMF was introduced in the context of the Extended
Press-Schechter formalism based on excursion-set theory \citep[hereafter
EPS,][]{Bond91, Bower91, Lacey93}.  Recent theoretical predictions use a
variant of the EPS formalism  in which the spherical collapse model is
replaced by an ellipsoidal collapse model \citep{Sheth02, Moreno08, Zhang08}.
An empirical fit to the CMF based on \nbs\ was presented by \citet{Cole08}.

Although the excursion-set models that employ ellipsoidal
collapse are successful in reproducing the overall (unconditional) halo mass
function as derived from \nbss\, these models do not provide accurate CMF
predictions.
Most of the current analytic predictions of the CMF deviate
from the results of \nbss\ by a multiplicative factor of a few,
especially for the number of massive progenitors. The empirical
fit of \citet{Cole08} does better, but its inaccuracies may
reach the level of 50\%. As shown by \citet{Cole08}, these inaccuracies
are valid whenever the standard variables are used, and they cannot be reduced
by a different functional fit.
In addition, this fit was calibrated against the Millennium simulation \citep{Springel05}
that was based on the cosmological parameters from the first-year data
of the Wilkinson Microwave Anisotropy Probe (WMAP),
it should be interesting to evaluate the CMF for the more up-to-date
cosmological parameters that emerge from the fifth-year WMAP data
\citep[e.g.][]{Komatsu09}.

The main goal of this paper is to provide a more accurate and more
robust empirical description of the CMF as measured from \nbss. We
work out the possible scaling laws which can be applied to the CMF
in order to capture its detailed properties over a large range of
cosmological models, halo masses and redshifts. The result is a
fitting function that offers a significant improvement in accuracy
over previous studies. The empirical fit can help us distinguish
between different analytical models of structure formation, and can
guide us to new improved versions of them. Such a fit can be used
for generating Monte-Carlo merger trees, which will accurately
reproduce the results of \nbss. The effect of cosmology, environment
density and different dark energy models can be studied, relating
these factors to haloes and galaxies \citep[e.g.][]{Maccio08}. These
issues and additional applications are discussed below in more
detail.

Currently, there is still some freedom in the definitions of halo
mass and merger trees. Accurate CMF fit may help us to test
different algorithms for constructing merger trees, including
assigning masses to haloes and relating progenitors to their
descendants. It will allow us to quantify the effects of different
construction schemes for merger trees, and especially to identify
the scheme that results in the simplest and most universal CMF.

A significant progress has been made in quantifying the overall halo mass
function, namely the abundance of haloes of a given mass at a given redshift.
The analytical model of \citet{Sheth99a} offers a significant
improvement over the classic estimate of \citet{Press74}.
Furthermore, the growing volume and dynamical range of \nbss\ allow
an accurate measurement of the halo mass function with small
sampling scatter and cosmic variance \citep[e.g.][]{Jenkins01,Warren06, Reed07}.
Still, there are significant deviations between the analytic approximations and
the simulations, which can increase as a function of redshift and reach a few tens
of percents at $\z=2.5$ \citep{Tinker08}.
For one thing, an accurate knowledge of the CMF can provide a better estimate
of the halo mass function and its evolution at high redshift.
However, the more interesting feature of the CMF is that it
involves much more information than the halo mass function, and it can
still be derived with a manageable statistical uncertainty.

Using the CMF, we present a new method for scaling a given set of
merger trees into a different cosmological model, mass resolution or redshift.
Although this technique requires an existing database of merger trees extracted
from a given \nbs, it offers an easy transformation to the desired cosmology,
mass and redshift. This method has certain advantages
over generating Monte-Carlo trees based on excursion-set models.
For example, it bypasses the inaccuracies associated with the
non-Markov features of the trees, associated with correlations between
adjacent time-steps \citep{Sheth04,Neistein08a}. This method
can be extended to include halo substructure and halo locations. Such a method
can be very useful for semi-analytical modeling of galaxy formation, black hole
growth, and dwarf galaxy assembly.

The common line of thought in quantifying merger trees starts with
the CMF. This is the basic prediction of the EPS
formalism, and it includes substantial information on the statistical
properties of the tree.
The known CMF allows derivations of other statistics
such as the main-progenitor history and merger rates.
However, \citet[][hereafter ND08a]{Neistein08a} have shown that an
alternative path of reasoning may also be useful, where the
main-progenitor history is derived first and a full
construction of merger trees follows. They found in \nbs\ that the
main-progenitor statistics is very regular and follows a log-normal
distribution of an appropriate mass variable, the variance $S$ of the
smoothed density field.
One of our purposes in the current paper is to verify this log-normal
behaviour in a range of cosmological models. We hope that this
will bring more insight into the physical origin of the main-progenitor
statistics.

It should be noted that the CMF does not include all the information
needed for describing merger trees. There are in general many
different subsets of trees that can accurately fit a given CMF.
Nonetheless, we will show in this work that the scaling laws of the CMF
can provide a good estimate for the full statistics of the
merger trees, including main-progenitor histories and merger rates.

This paper is organized as follows. In \S\ref{sec:sims} we describe
the suite of \nbss\ used and the way merger trees are constructed.
Section \ref{sec:cmf} is devoted to the conditional mass function,
where we study its scaling properties and provide an accurate
fitting function. In \S\ref{sec:mp} we discuss an alternative
description, using the log-normal nature of the main-progenitor history.
A simple prescription
on how to scale a given simulation is developed in \S\ref{sec:appl}.
We summarize the results and discuss them in \S\ref{sec:discuss}.
Additional statistical properties of merger trees which are not
critical for the body of the paper are added in Appendix
\ref{sec:add_stat}. Throughout the paper we use $\log$ for
$\log_{10}$ and $\ln$ for natural logarithm.


\section{The Simulations}
\label{sec:sims}

All simulations have been performed  with \textsc{pkdgrav}, a tree code written
by Joachim Stadel and Thomas Quinn \citep{Stadel01}. The code uses spline
kernel softening, for which  the forces become completely Newtonian at
2  softening lengths.   Individual time  steps for  each  particle are
chosen  proportional  to the  square  root  of  the softening  length,
$\epsilon$,    over   the   acceleration,    $a$:   $\Delta    t_i   =
\eta\sqrt{\epsilon/a_i}$. Throughout, we set $\eta = 0.2$, and we keep
the  value of the  softening length  constant in  co-moving coordinates
during each run ($\epsilon$=1.62 $h^{-1}$kpc).
Forces  are  computed using  terms  up  to
hexadecapole order  and a node-opening angle $\theta$  which we change
from $0.55$ initially  to $0.7$ at $\z=2$.  This  allows a higher force
accuracy when the mass distribution  is nearly smooth and the relative
force errors can  be large. The initial conditions  are generated with
the \textsc{grafic2} package \citep{Bertschinger01}.  The starting redshifts $\z_i$
are  set to  the  time when  the  standard deviation  of the  smallest
density fluctuations resolved within  the simulation box reaches $0.2$
(the smallest scale resolved  within the initial conditions is defined
as twice the intra-particle distance).
For each simulation we stored more than 100 outputs from redshift 10
to redshift zero, in order to construct detailed merger trees.
The parameters of the simulations used in this work are describe in Table \ref{tab:sims}.

\begin{table}
\caption{A summary of the \nbss\ used in this work, all with flat
cosmology and with a Hubble constant of 72 km/s/Mpc. Particle mass
is in units of $\hmsun$, box size is in Mpc. Each simulation follows
the evolution $350^3$ particles.}
\begin{center}
\begin{tabular}{lcccccc}
\hline Name & $\Omega_m$ &  $\sigma_8$ & Particle mass & Box \\
\hline
wmap5 & $0.258$ &  $0.796$ &  $4.54\times10^8$ & 90  \\
lcdm & $0.258$ &  $0.915$ & $4.54\times10^8$ & 90  \\
scdm1 & $1.0$ & $0.77$ & $1.76\times10^9$ & 90 \\
scdm2 ($n=-2$) & $1.0$ & $0.8$  & $1.76\times10^9$ & 90 \\
\hline
\end{tabular}
\end{center}
\label{tab:sims}
\end{table}

In  all  simulations,  dark  matter  haloes  are  identified  using
the \textsc{fof} algorithm \citep{Davis85} with linking length of 0.2 times the mean
interparticle separation. Only haloes which include more than 20
particles are saved for further processing. For constructing merger
trees, we started marking all the particles within the virial radius
of a given haloes at $\z=0$ and we tracked them back to the previous
output time. We then make a list of all haloes at that earlier
output time containing marked particles, recording the number of
marked particles contained in each one. In addition we record the
number of particles that are not in any halo in the previous output
time and we consider them as {\it smoothly} accreted.

We used the two criteria suggested in  \citet{Wechsler02} for halo 1 at one output
time to be labeled a ``progenitor'' of halo 2 at the subsequent output
time.  In our language halo 2 will then be labeled as a ``descendant''
of halo 1 if i) more than 50\% of the particles in halo 1 end up
in halo  2 or  if ii) more than 75\% of halo 1 particles that end
up in any halo at time step 2, do end up in halo 2 (this second
criterion is mainly relevant during major mergers).
Thus a halo can have only one descendant but there is no limit to the
number of progenitors.

We found evidence for the so-called `backsplash' subhalo population
\citep[e.g.][]{Knebe08}. These haloes have orbits that brought them
inside the virial radius of their host at some earlier time, but
without them been really accreted (i.e. they managed to come out
from their host dark matter halo).
We decided to treat them in two different ways according to their
final fate: i) If after have been inside the main halo, the
backsplash halo survives as isolated halo till the present time,
than it is removed from the progenitor list of the parent halo (i.e.
it is removed from the merger tree); ii) if the halo is accreted
again (in a definite way) by the main halo at a later time step then
it is considered as accreted the first time it entered the main
halo. We found that back splash haloes are roughly 8\% of the total
progenitor number but they only marginally contribute (less than
2\%) to the final halo mass.

\begin{figure}
\centerline{ \hbox{ \epsfig{file=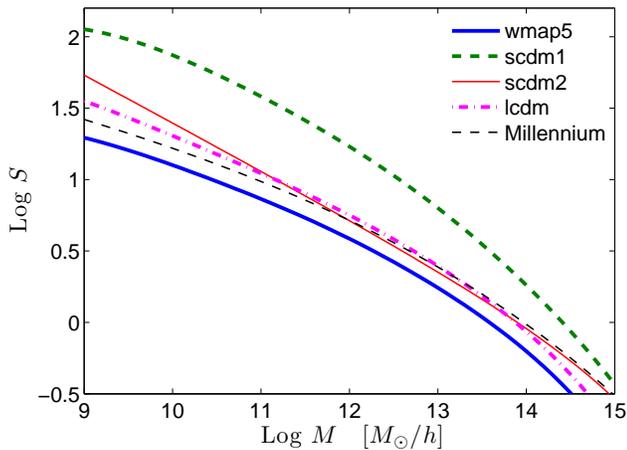,width=9cm} }}
\caption{The power spectra of the density fluctuation fields in the simulations.
Shown is $S(M)=\sigma^2(M)$, the variance of the initial mean density
in spheres that encompass mass $M$, linearly extrapolated to $\z=0$.
The Millennium simulation \citep{Springel05} is shown for reference.}
  \label{fig:various_s}
\end{figure}

According to the Extended Press-Schechter formalism the statistical
properties of merger trees are fixed by the density field at early
times when perturbations grow linearly. All the statistical
properties of this field are described by its variance,
$S(M)=\sigma^2(M)$. Specifically, $S(M)$ is the variance of the
density field, smoothed with a spherical top hat filter in space,
and linearly extrapolated to $\z=0$ \citep[for more details,
see][]{Lacey93}. In fig.~\ref{fig:various_s} we plot $S(M)$ for all
the simulations used in this work. This comparison shows the
predicted similarity between merger trees of different simulations:
we have two simulations with very similar $S(M)$ but different $\omm$
(lcdm \& scdm2). One simulation has a low value of $\sigma_8$
(wmap5), and the scdm1 simulation has a very different shape of
$S(M)$.

In order to avoid deviations of $S(M)$ due to the small box size and
cosmic variance we measured it directly from the initial condition
density field used for each simulation. The values of $\sigma_8$
given in table \ref{tab:sims} are those obtained by this calibration
method. For the scale free
simulation (scdm2) the box size is important as it limits contributions from
large scales to the field variance \citep[see a detailed discussion in][]{Smith03}.
This effect bends $S(M)$ slightly at the high mass
end, in agreement with the theoretical prediction.


\section{Conditional Mass Function}
\label{sec:cmf}

This section is devoted to a comparison of the conditional mass
function (CMF) as extracted from our suite of \nbss. The scaling laws
of the CMF are important as they highly constrain the behaviour of
the full statistics of merger trees. In Appendix \ref{sec:add_stat}
we show other statistical properties of trees, and demonstrate that
they follow approximately the properties of the CMF found here.

According to the Extended Press-Schechter formalism
\citep[EPS,][]{Bond91,Bower91,Lacey93}, the average number of
progenitors in the mass interval $[M, M+\dd M]$, which will merge
into a descendant halo $M_0$ after a timestep $\dW$, is given by
\begin{eqnarray}
\label{eq:dNdM}
\frac{{\dd}N}{{\dd}M} (M | M_0,\dW) \dd M \,
\; = {M_0 \over M} f(\dS,\dW) \left\vert
\frac{\dd S}{\dd M} \right\vert \dd M\;   \,.
\end{eqnarray}
Here $\omega \equiv \delta_{c}(\z)/D(\z)$, where $\delta_c(\z)\simeq
1.68$ with a weak dependence on $\z$, and $D(\z)$ is the
cosmological linear growth rate. $\dW\equiv \omega-\omega_0$ where
the progenitors are identified at $\omega(\z)$ and the descendant
halo at $\omega_0$. We refer the reader to the appendix of ND08a for
a detailed summary on how to compute $\omega$, and a simple fitting
function. The quantity $f\,\dd S$  describes the fraction of mass
of the descendant halo that is in progenitors of
mass in the range $[S, S+\dd S]$. According to EPS
\begin{eqnarray}
\label{eq:m_frac}
f_{\rm ps}(\dS,\dW) = \frac{1}{\sqrt{2 \pi}} \; {\dW \over (\dS)^{3/2}}
\; {\rm exp}\left[-{(\dW)^2 \over 2 \dS}\right] \,,
\end{eqnarray}
where $\dS\equiv S(M)-S(M_0)$,
and $f_{\rm ps}$ is the specific solution of $f$ given by the EPS
formalism. In what follows we will use the
name `CMF' to designate $\dd N/\dd M$ only.

In the language of the EPS formalism $f_{\rm ps}$ is `universal' -- it does not explicitly depend
on the descendant halo mass $M_0$ or on the background cosmology.
These parameters appear only in the transformation of
$\dS$ and $\dW$ to mass and time (or redshift).
The limited success of the EPS formalism in reproducing the \nbs\ merger trees
indicates that the function $f_{\rm ps}$ is a limited approximation,
but it is possible in principle that $f_{\rm ps}$ has universal properties.
Both \citet{Sheth02} and \citet{Cole08} found that $f$ as derived from
\nbss\ shows non-negligible deviations from universality.
However, these studies did not explore in detail
the breakdown of universality and did not test other models for $f$
against \nbss. For example, \citet{Cole08} have studied $f$ only for
descendant haloes identified at $\z=0$ and showed deviations of
$\sim50\%$ for various values of $\dW$ and halo masses.

The universality of $f_{\rm ps}$ is based on the two main theoretical elements of the EPS formalism,
namely the dynamics of the spherical collapse model and the statistical properties of the initial
density fluctuation field when smoothed with a top-hat filter in $k$-space.
Variants of these elements have been attempted, including ellipsoidal collapse
or smoothing filters of different shapes
\citep{Bond91,Sheth02,Zentner07,Desjacques08}. In these models
$f$ may in principle depend on the descendant halo mass and thus become non-universal,
but this is not necessarily so.
For example, \citet{Moreno08} showed that when ellipsoidal collapse is assumed, $f$ remains
universal if the variables $S(M)$ and $\dW$ are normalized by $S(M_0)$ and
$\sqrt{S(M_0)}$ respectively.
In its most general form, $f$ may depend on $S_0\equiv S(M_0)$,
$\omega_0$ and the specific cosmology used. We will show below that
neglecting the dependence on $\omega_0$ and on cosmology may not be too harmful,
but the dependence on $S_0$ must be properly taken into account for an accurate
description of the CMF.


\subsection{Self-similarity in time}
\label{sec:cmf_time}

According to eq.~\ref{eq:dNdM}, the CMF as predicted by EPS depends
only on $\dW$ and not on the redshift $\z_0$ where the descendant
halo is defined. This self-similarity implies that merger trees
extracted at different redshifts are self-similar when using
$\omega$ as the time variable. Indeed, ND08a and \citet{Genel08}
have verified this behaviour using merger trees extracted from the
large Millennium simulation \citep{Springel05}. It was shown to work
well for main-progenitor histories and merger rates, with scatter of
a few percent up to $\z\sim5$ \citep[see also][who found deviations
from this symmetry using a different definition of the halo
mass]{Fakhouri08}.

\begin{figure}
\centerline{ \hbox{ \epsfig{file=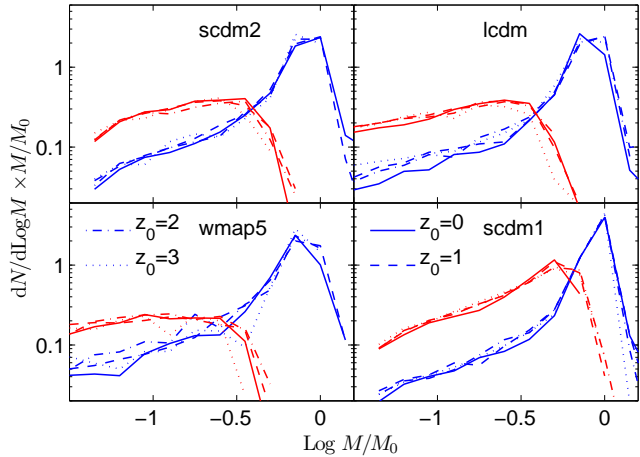,width=9cm} }}
\caption{Self-similarity in time of the progenitor conditional mass function (CMF).
Each panel shows the CMF as measured from a different \nbs\ as indicated. The descendant
halo is identified at different times corresponding to redshifts $\z_0=0,1,2,3$
(solid, dashed, dotted-dashed, and dotted
lines respectively). The progenitors are identified at two lookback times
corresponding to $\dW=0.5,3$ (blue and red lines respectively).
The distribution of descendant masses is selected in the range $10^{12}\leq M_0 \leq 10^{13}\,\hmsun$
such that it produces the same distribution of $M_0$ in the different cases of $\z_0$.
In each case, the four curves show only a weak dependence on the descendant time $\z_0$.
}
  \label{fig:cmf_time1}
\end{figure}

In fig.~\ref{fig:cmf_time1} we show the level of self-similarity in
time in our four \nbss. The scdm1 and scdm2 simulations show
small variations in the CMF between different values of $\z_0$, consistent
with the sampling noise. The wmap5 and lcdm cosmologies show larger
deviations, up to a factor of two for small progenitors and for the
time-step $\dW=0.5$. However, the data do not show a monotonic
trend with $\z_0$, hinting that our small-number statistics may contribute
to the scatter. For example, the number of descendant haloes within
a mass range $10^{12}\leq M_0 \leq 10^{13}\,\hmsun$ in the wmap5
cosmology is roughly (500,400,100,100) for $\z_0=(0,1,2,3)$
respectively. It is encouraging that better results were
obtained for the Millennium simulation (as described above) and for
large time-steps. Nonetheless, part of the deviation from self-symmetry may
be related to the way haloes are defined using \textsc{fof}
linking length and the variation with redshift implied by this definition.
As explained in section \ref{sec:sims}, we try to correct the merger trees for
'backsplash' haloes. This correction may introduce some additional
asymmetry between $\z_0=0$ and higher redshifts.

\begin{figure}
\centerline{ \hbox{ \epsfig{file=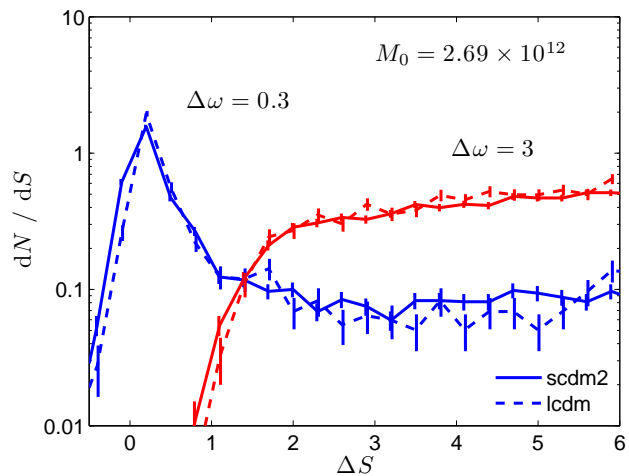,width=9cm} }}
\caption{Self-similarity of the CMF with respect to the cosmological expansion history.
The two sets of CMF are computed for $\z_0=0$, for descendant mass in the range
$10^{12}\leq M_0 \leq 10^{13}\,\hmsun$, and for $\dW=0.3,3$
(blue and red lines respectively). The solid and dashed lines refer
to the scdm2 and lcdm simulations respectively.
The distribution of descendant masses at $\z_0$ is matched in order to produce
the same distribution of $S(M_0)$ values, and the average descendant halo mass is indicated
in units of $\hmsun$. Note that we plot the number of progenitors per unit of $S$, which
is given by eq.~\ref{eq:dNdM} times $|\dd M/\dd S|$.
The CMF shows only a weak dependence on cosmology.}
  \label{fig:cmf_cosmo}
\end{figure}

The fact that merger trees remain similar when the descendant time is varied
from $\z=0$ to $\z=3$ is a non-trivial result.
The different values of the cosmological parameters
$\omm$ and $\oml$ at high redshift imply that the merger trees for a higher $\z_0$
evolve in a different cosmological background. Thus, the dependence
of merger trees on the cosmological parameters $\omm$ and $\oml$ should be
all folded into the time variable $\omega=\delta_c/D$. In
fig.~\ref{fig:cmf_cosmo} we test this hypothesis explicitly by comparing the
CMF from the scdm2 and lcdm simulations. These simulations have a
similar shape of $S(M)$ (see fig.~\ref{fig:various_s}) but very
different $\omm$ values. The agreement is within the Poisson
sampling error-bars, demonstrating that $\omega$ can scale properly
different expansion histories of the Universe.

\subsection{Different power-spectrum}
\label{sec:cmf_ps}

\begin{figure}
\centerline{ \hbox{ \epsfig{file=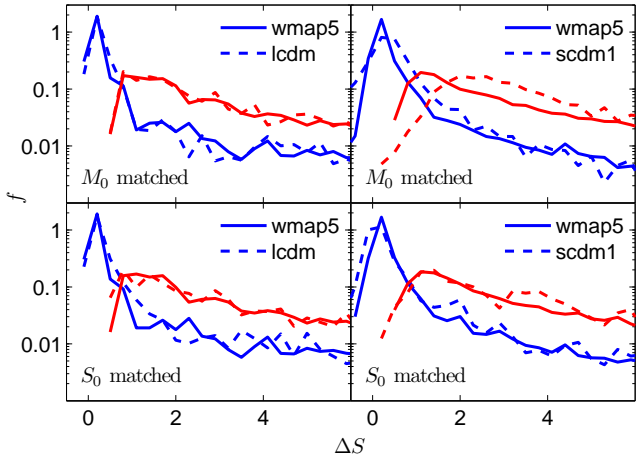,width=9cm} }}
\caption{Fraction of mass in progenitors for different power spectra and
cosmologies.
The lines of two types in each panel compare results from two different
simulations as indicated.
The time steps are $\dW=0.3,2$ and the descendant haloes are selected at
$\z_0=0$.
In the left and right panels the descendant halo masses in wmap5 were
selected in the range $10^{13}\leq M_0 \leq 10^{14}\,\hmsun$ and
$10^{12}\leq M_0 \leq 10^{13}\,\hmsun$ respectively.
In lcdm and scdm1 the descendant haloes were chosen to match the distribution
of $M_0$ or $S_0$ in wmap5, as indicated in each panel.}
  \label{fig:cmf_ps}
\end{figure}

Following the EPS formalism (eq.~\ref{eq:m_frac}) the fraction of
mass inside progenitors, $f_{\rm ps}(\dS,\dW)$, depends only on
$\dW$ and $\dS$. As mentioned above, one could assume less universal
form, $f(\dS,\dW|S_0)$ that will enable accurate description of the
CMF for various cosmological models. In order to test this we plot
$f$ in fig.~\ref{fig:cmf_ps} for two sets of power-spectra. The
results of the first comparison (left panels) between lcdm and wmap5
simulations is very good, showing no significant deviation in $f$
between these models. This is obtained when the same values of $M_0$
are selected in each cosmology, or the same $S_0 = S(M_0)$. The
difference between these two selection criteria is negligible, so we
actually sample $f(\dS,\dW|S_0)$ for a small range in $S_0$.

The second comparison (right panels) shown in fig.~\ref{fig:cmf_ps}
is for wmap5 and scdm1 cosmologies. As seen in
fig.~\ref{fig:various_s} both $S(M)$ and its derivative $\dd S/\dd
M$ vary significantly between these two simulations. This is being
translated into a large discrepancy in $f$ when the same values of
$M_0$ are chosen from both simulations. However, selecting a sample
with the same $S_0$ values leaves $f$ invariant, proving that
$f=f(\dS,\dW|S_0)$ (at least for our limited set of data). For a
given $S_0$ the values of $M_0$ in both simulations differ by a
factor of $\sim$10, limiting the dynamical range for which we can
check this $f$ scaling.

The dependence of $f$ on $S_0$ deserves a more careful test, probably with a larger set
of \nbss, spanning a larger range of power-spectra. It is important to verify that such
a dependence is not related to the \emph{shape} of $S(M)$, as was found here.
If this is true, it implies that a modification to the normalization
of the power-spectrum ($\sigma_8$) which changes $S_0$, will induce
a non-trivial change to the CMF. Such a dependence on $\sigma_8$ is
different from the simple scaling of the shape $\dW/\sqrt{\dS}$ that
is used by EPS (see below).


\subsection{Global fitting function}
\label{sec:cmf_fit}

\begin{figure}
\centerline{ \hbox{ \epsfig{file=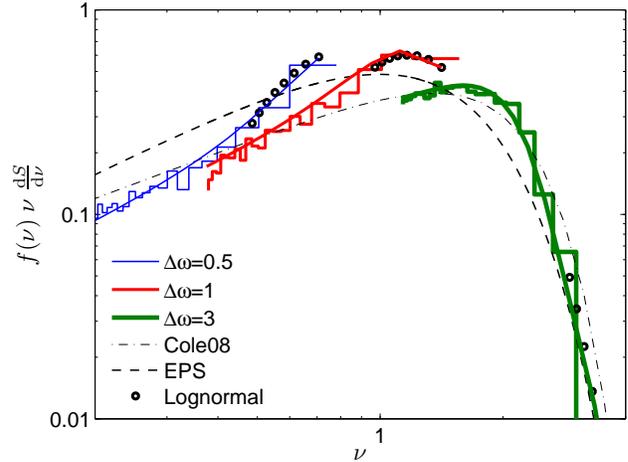,width=9cm} }}
\caption{The limitation of using a fitting function of the form $f(\nu|S_0)$.
The results form the scdm2 simulation are plotted as histograms.
They are extracted at $\z_0=0$ with three different time steps $\dW=0.5,\,1,\,3$
and for an average halo mass of $3\times10^{12}\,\hmsun$.
The global fit of eq.~\ref{eq:cmf_fit} for the three time steps is shown as smoothed solid curves.
The EPS prediction is plotted as a thin dashed line
and the fit of \citet{Cole08} is shown as a dotted-dashed line.
Like any other function of the type $f(\nu)$, they both fail to reproduce the simulation results.
The lognormal global fit to $P_1$ (eqs.~\ref{eq:p1_lognormal} \& \ref{eq:p1_global})
is plotted as small circles (only
for $M>M_0/2$, where it should be identical to the CMF, see section \ref{sec:mp}).
}
  \label{fig:cmf_fnu}
\end{figure}

The EPS formalism suggests that the mass ($\dS$) and time ($\dW$)
variables can be scaled into a new variable
$\nu\equiv\dW/\sqrt{\dS}$. In terms of $\nu$, the function
$f(\nu)\,\nu \, \dd S /\dd \nu$ should be universal. As shown above
some dependence on $S_0$ is required, so we can try to use a
function of the shape $f(\nu|S_0)$. This is tested in
fig.~\ref{fig:cmf_fnu} where we plot $f(\nu|S_0)$ for a fixed value
of $S_0$ and for various $\dW$. The clear deviations from a unique
line show that such a formulation is too simple and incompatible
with the results of \nbss. This figure is very similar to the one
given in \citet{Sheth02} and \citet{Cole08}, and agrees with these
previous results. Thus, we are forced to use a more general form
$f=f(\dS,\dW|S_0)$.

We found that the fitting function below can fit the CMF from our
set of \nbss\ for all time-steps larger than $\dW\sim0.5$ and all
$S_0$ values. We have looked for the simplest possible function that
can fit the data, which is still similar to the EPS original
function. The function we adopt is:
\begin{eqnarray}
\label{eq:cmf_fit}
\lefteqn{ f(\dS,\dW|S_0) = a \frac{\dW}{\dS^{1.5}} \exp \left[ -b\frac{\dW^2}{\dS} \right] + } \\ \nonumber & &
c \frac{\dW^2}{\dS^3} \exp \left[ -d \frac{\dW^{2.5}}{\dS^{2+\alpha}} \right] \, ,
\end{eqnarray}
where the parameters explicitly depend on $S_0$ in the following way:
%
%
\begin{eqnarray}
\label{eq:cmf_prm}
\nonumber a &=& 0.215 + 0.0037\, S_0 \\
\nonumber b &=& 0.414 + 0.0013\, S_0^2  \\
c &=& 0.0746 + 0.0382 \, S_0 \\
\nonumber d &=& 0.257 + 0.0568 \, S_0 \\
\nonumber \alpha &=&  - 0.0141 \, S_0 + 0.0056 \, S_0^2
\end{eqnarray}
The added term is not a natural feature of the standard EPS
formalism, and it emphasizes the need for additional physics, beyond
the Sheth-Tormen generalization from spherical to ellipsoidal
collapse model. The failure of functions that depend solely on $\nu$
in reproducing the detailed evolution of \nbss\ may have interesting
theoretical implications. We also note that our fitting function is
not necessarily unique --- there may exist other fitting functions
of comparable quality. However, it is likely to be sufficient for
most applications related to galaxy formation, based on the tests
performed here over wide ranges of halo masses, redshifts and
cosmological models.

\begin{figure}
\centerline{ \hbox{ \epsfig{file=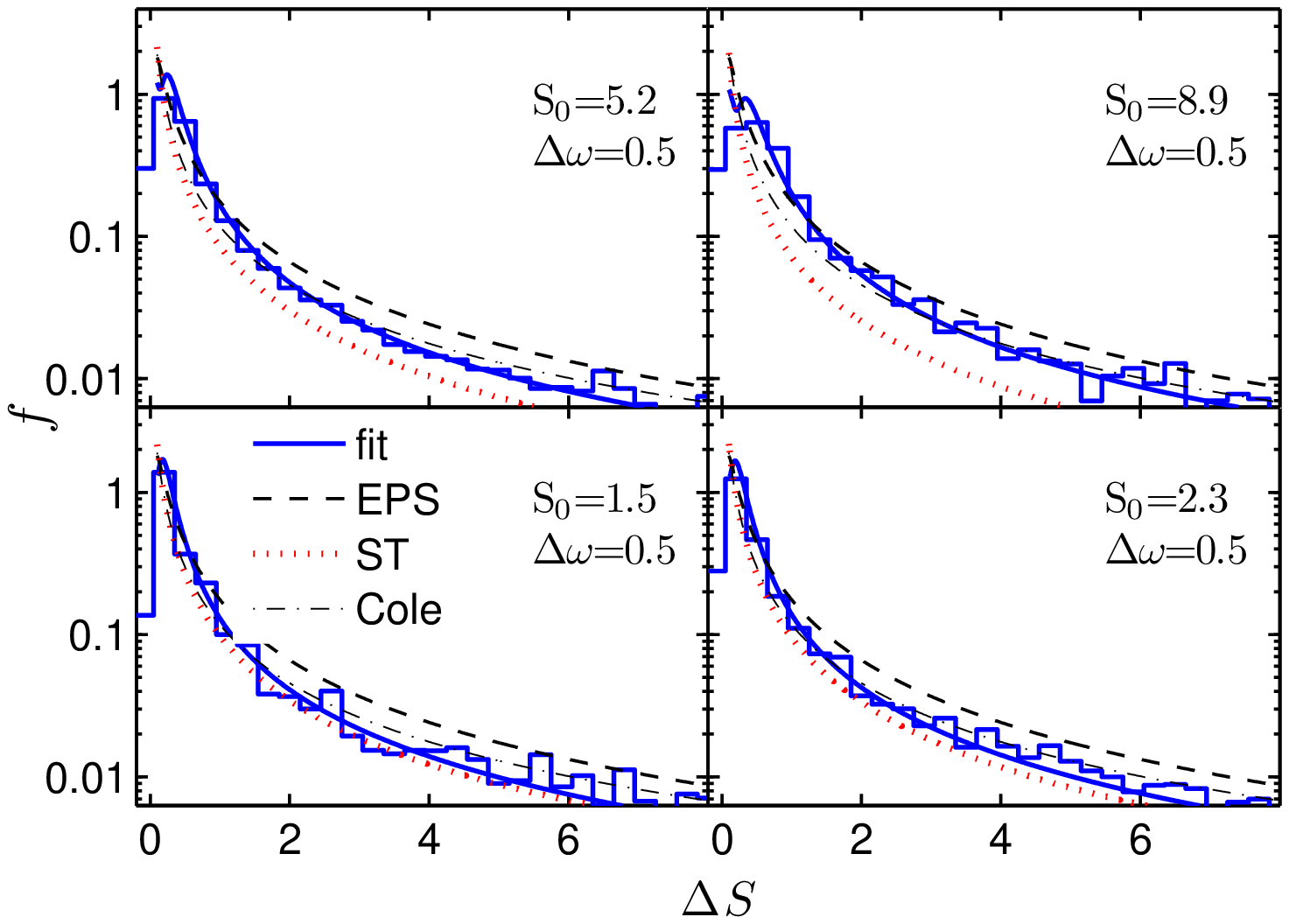,width=9cm} }}
\centerline{ \hbox{ \epsfig{file=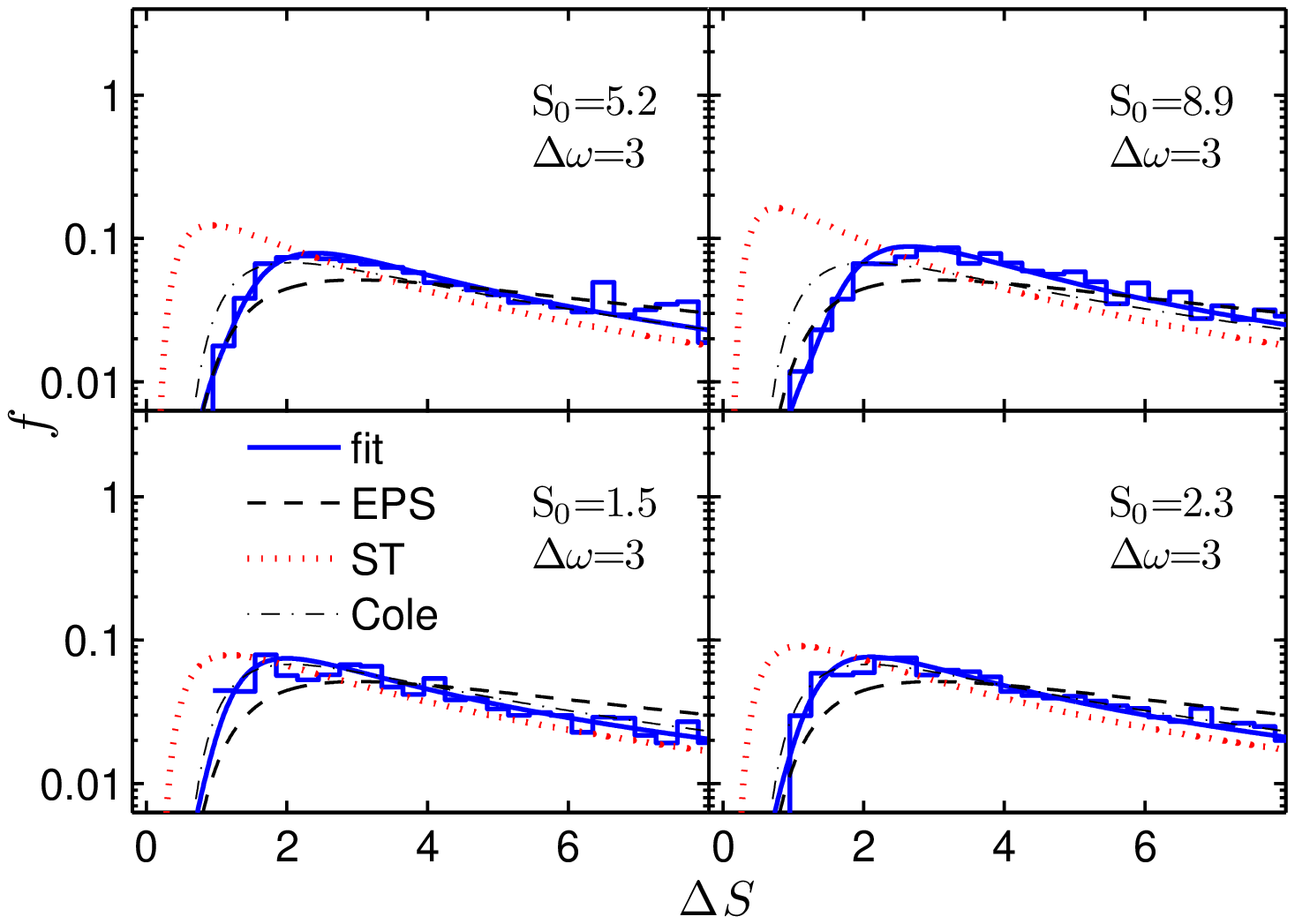,width=9cm} }}
\caption{A global fit to $f(\dS,\dW|S_0)$ as
defined in eqs.~\ref{eq:cmf_fit} and \ref{eq:cmf_prm}. Each panel
refers to a given set of $S_0$ and $\dW$ values as indicated.
The simulation data (histograms) were extracted and averaged from all the different simulations
used here at $\z_0=0$ and at $\z_0= 1$. The descendant masses were selected in log bins of
size half dex.
Our global fit, the EPS prediction, the fit of \citet{Cole08},
and the fit of \citet{Sheth02} are shown as solid, dashed, dashed-dotted and dotted lines
respectively.
Note that both the EPS prediction and the fit of \citet{Cole08} are independent of $S_0$.}
  \label{fig:cmf_fit}
\end{figure}

In fig.~\ref{fig:cmf_fit} we compare our fitting function to the
simulation data, showing that it accurately reproduces the trends
found. The agreement between the fit and data is at the level of the
statistical noise. The dependence on $S_0$ is mainly seen at low
$\dS$ (corresponding to massive progenitors). We also show the EPS
prediction and the fit derived by \citet{Cole08}, both independent
on $S_0$. Evidently, our fit agrees well with \citet{Cole08} for
intermediate and massive haloes (although deviations for massive
progenitors are highlighted in fig.~\ref{fig:cmf_fnu}). This is
encouraging because the fits are based on different simulations and
different merger-tree construction schemes. However, our fit breaks
the symmetry of using only $\nu$ as was done by \citet{Cole08}, in
this way it can capture the behaviour of the data for low mass
haloes, and across various time-steps. The accuracy for different
$\dW$ can be specifically seen in fig.~\ref{fig:cmf_fnu}.

The integral of $f$ can be computed analytically only for $\alpha=2$ ($S_0\sim 2.5$),
for the whole halo mass range we computed the integral numerically, yielding 0.75 up to 0.9,
depending on the descendant mass and on the time-step.
This predicts that a substantial fraction of the mass is not included in any progenitors but rather
accreted from a  `smooth' component.

Throughout this work we mainly discuss the EPS formalism in its
standard version. However, as mentioned in the introduction,
versions of this formalism which use the ellipsoidal collapse model
are presumably more accurate, as is indicated by their ability to
predict accurate halo mass functions. We examine two such studies,
the pioneer work of \citet{Sheth02} and a more recent work by
\citet{Moreno08}. Our interpretation of both studies is that the CMF
as predicted by the spherical model is comparable in its accuracy to
the one predicted by ellipsoidal collapse. This can be seen in figs.
7 \& 8 of \citet{Sheth02}. More evidently, figs.~3-5 of
\citet{Moreno08} show that the spherical model gives better results
for low mass descendant haloes, but this trend changes for massive
haloes, where the ellipsoidal collapse is more accurate. We also
show in fig.~\ref{fig:cmf_fit} the CMF proposed by \citet{Sheth02}
\footnote{We adopt eq.~7 from \citet{Sheth02} and use $n!$ in the
denominator of $T$ instead of $n$. According to J. Moreno (private
communications) this seems to be a typo.}. It is clear that this
model predicts CMFs which are significantly higher from the
simulation results in the regime of massive progenitors and large
time-steps. Note that plotting $f(\dS,\dW|S_0)$ shows more clearly
this effect than the usual plot as a function of mass ratio
\citep[e.g.][]{Moreno08}. In Appendix \ref{sec:app_cmf} we provide
more results in the form of fig.~\ref{fig:cmf_fit}


\section{Main-Progenitor Histories}
\label{sec:mp}

The ``main-progenitor'' history of a given merger-tree is
constructed by following backward in time the most massive progenitor in
each merger event. This is a useful definition as it allows us to
follow a well defined branch of the tree. In addition, a
quantitative description of the main-progenitor history highly
constrains the full statistics of trees.
Properties of main-progenitor histories were studied extensively,
both analytically and using \nbss\ \citep{Lacey93,
Nusser99,Firmani00, Wechsler02,vdBosch02b, Neistein06, Li07,
Neistein08a, Zhao08}. Let us also mention that the main-progenitor
is not always the most massive  progenitor at a given time.

We define $P_1$ as the probability density to find a main-progenitor of mass $S(M_1)$
for a given descendant halo mass $M_0$, and a time-step $\dW$. As was found by ND08a,
$P_1$ can be well fitted by a lognormal distribution,
\begin{equation}
\label{eq:p1_lognormal}
P_1(\dS_1|S_0,\dW) = \frac{1}{\sigma_p \dS_1 \sqrt{2\pi}} \exp \left[ - \frac{
(\ln\dS_1-\mu_p)^2}{2\sigma_p^2} \right] \, .
\end{equation}
Here $\dS_1=S(M_1)-S(M_0)$, the parameters $(\sigma_p, \mu_p)$ depend on $S_0$ and $\dW$,
and as usual $S_0\equiv S(M_0)$.
By definition, the integral over $P_1(\dS_1)$ equals unity for any $S_0$ and $\dW$.

\begin{figure}
\centerline{ \hbox{ \epsfig{file=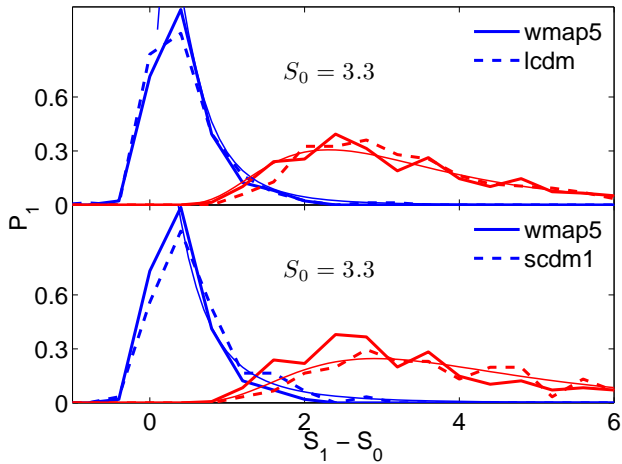,width=9cm} }}
\caption{The distribution of main progenitor masses in different cosmological
models. Results are shown for descendant haloes defined at $\z_0=1$ and for main progenitors
at two different time steps, $\dW=0.5,\, 3$. The descendant halo mass is in the range $10^{12}
\leq M_0 \leq 4\times 10^{12} \,\hmsun$ for the wmap5 cosmology, and $M_0$ in the other cosmologies
were selected to match the same distribution of $S(M_0)$. Thin solid lines show the
lognormal global fit, eq.~\ref{eq:p1_global}, as computed for the lcdm and scdm1 cases.}
  \label{fig:P1_cosmo}
\end{figure}

Main-progenitor histories were constructed for all the simulations used
in this work. Following ND08a, we confirm that the lognormal distribution
fits accurately the simulation results. This was tested for all possible
ranges of halo masses, redshift, and time-steps. As was mentioned by ND08a the fit
becomes inaccurate for small time steps, typically for $\dW\lesssim 0.5$. Such a
behaviour is found here as well. One exception to the above, where the lognormal fit
is somewhat innacurate, is for the scdm1 simulation at $\z_0=0$. This may be due to the
high sensitivity of $\dS$ in this cosmological
model ($\dd S/\dd M$ is typically 7 times larger than other cases). The fact that the fit
works well for $\z\gtrsim 1$ data using the same cosmological simulation is encouraging.
In fig.~\ref{fig:P1_cosmo} we plot main-progenitor distributions for few cases of halo
mass, $\z_0$ and $\dW$, more examples can be found in Appendix \ref{sec:app_mp}. The accuracy of the lognormal
fit as plotted here is typical for the rest of the cases.

For each cosmology we are able to use a global fit, similar to the
one suggested by ND08a\footnote{Here we parameterize the fit
according to $S_0$ in order to stress universal features, and not
according to $M_0$ as was done in ND08a}. The fit approximates the
standard deviation and average of $\ln \dS_1$ by
\begin{eqnarray}
\label{eq:p1_global}
\sigma_p &=& (a_1\log S_0+a_2) \log \dW + a_3\log S_0 + a_4 \\ \nonumber
\mu_p &=& (b_1\log S_0+b_2) \log \dW + b_3\log S_0 + b_4 \,.
\end{eqnarray}
We provide numerical details for the values of these parameters in
Appendix \ref{sec:app_mp}. For a given $S_0$, the evolution with
time of $\sigma_p$ and $\mu_p$ is simply linear in $\log \dW$. In
fig.~\ref{fig:cmf_fnu} we show that this evolution gives a very
accurate fit to the relevant part of the CMF (the CMF and
main-progenitor distribution are identical for $M>M_0/2$). We recall
that such time-evolution was obtained for the CMF using much more
complicated $\dW$ dependence (see eq.~\ref{eq:cmf_fit}). Although
both approaches discuss only the fitting possibilities, it seems
that the main-progenitor gives a much simpler way to describe the
merger history in an accurate way.

The simple behaviour of $P_1$ for each cosmology as seen in eq.~\ref{eq:p1_global} and in
fig.~\ref{fig:cmf_fnu} calls for a more global fit, which predicts the parameters of $P_1$
in other cosmological models. We compared
$P_1$ for our set of simulations and found a high level of self-similarity,
with the exception of the scdm1 case. Unfortunately  we could not find a general
law able to combine scdm1 with the other cosmological models. For example,
choosing haloes with the same values of $S_0$ from scdm1 and other simulations
does not yield the same results.
To summarize, our study indicates that the main-progenitor distribution as extracted from
\nbss\ has a universal lognormal shape. However, the dependence of the lognormal parameters
on cosmology is not clear, and there is no theoretical explanation
to this phenomena yet.

Recently, \citet{Zhao08} studied the behaviour of main-progenitor
histories for different cosmological models. They provide a fitting
procedure for estimating \emph{median} main-progenitor histories in
any cosmological model. However, we find it hard to estimate the
full distribution $P_1(\dS)$ out of their algorithm.

\section{Applications}
\label{sec:appl}

An accurate fitting function for the CMF includes substantial
information on halo growth, and it has various interesting
applications: it can shed more light on the evolution of the halo
mass function with redshift and cosmology \citep[e.g.][]{Tinker08},
it can constrain dark energy models, it is useful for generating
Monte-Carlo merger trees, and for predicting mass-accretion
histories of the main-progenitor. Here we focus on a new methodology
for re-scaling a given set of merger trees between different
cosmological models. We also discuss briefly the possibility of
generating Monte-Carlo trees.

\begin{figure}
\centerline{ \hbox{ \epsfig{file=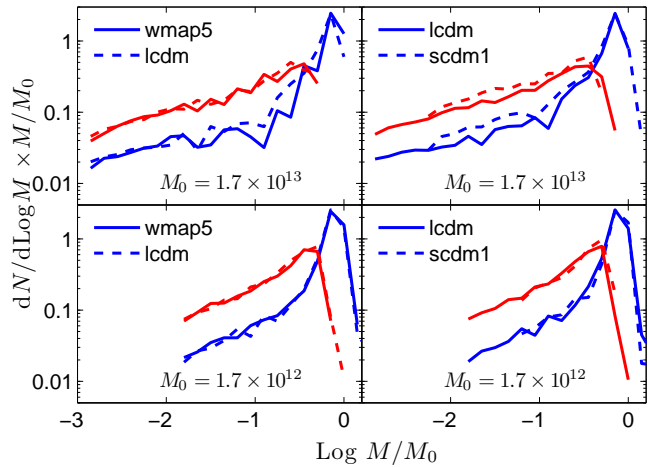,width=9cm} }}
\caption{The conditional mass function (CMF), $\dd N/\dd M$, computed for different cosmological
models using a transformation of the time-step $\dW$.
In each panel we compare results from two different
simulations and for the same descendant halo mass as indicated (in units of $\hmsun$).
The time steps used for the lcdm cosmology are $\dW=0.5,\, 2$.
For the wmap5 and scdm1 simulations they are different by a factor of 0.86 and 1.8 respectively.
We learn that the time transformation allows a reasonable scaling of the merger trees,
albeit with certain deviations in some cases.}
  \label{fig:cmf_scaling}
\end{figure}


\subsection{Re-scaling a given merger tree}
\label{sec:rescale}

There are many existing resources of \nbss\ and related merger trees
which are publicly available. The most useful one being the
Millennium simulation \citep{Springel05} with its web-based database
\citep{Lemson06}. However, recent changes in the observed values of
cosmological parameters (especially $\sigma_8$) make these
simulations inaccurate for predicting observable quantities in our
Universe. Here we suggest a new methodology to transform merger
trees into a different cosmological model, halo mass, and redshift.
Such transformation can also be useful for enhancing the mass
resolution of merger trees. There are few benefits for this approach
over the standard method of generating Monte-Carlo trees: (a) it can
preserve the non-Markov behaviour of trees (see e.g. ND08a) (b) it
might be easily extended to handle substructures (c) it might be
extended to accurate transformation of halo spatial locations.

We define the case $r$ as our reference data from a given merger-tree,
\begin{equation}
\label{eq:ref_case}
\{M_i \, | M_0, \dW, C \}_r \,
\end{equation}
The data is defined as a set of progenitors $M_i$ for a given descendant halo mass $M_0$,
time-step $\dW$ and a cosmological model $C$. Our target case is defined as a different cosmology
$\widetilde{C}$ and descendant mass $\widetilde{M}_0$
\begin{equation}
\label{eq:tar_case}
\{ \widetilde{M}_i \, | \widetilde{M}_0, \widetilde{\dW}, \widetilde{C} \}_t \ .
\end{equation}
We are looking for a transformation of the kind
\begin{equation}
M_i \rightarrow \widetilde{M}_i \;\;\;\; , \;\;\;\; \dW \rightarrow \widetilde{\dW}
\end{equation}
that will yield a different set of progenitors, possibly for a
different time-step, which will be consistent with the target case.
Such a consistency is achieved when the transformed progenitors will
yield the proper CMF. Note that our global fit was done for
$f(\dS,\dW|S_0)$, as seen from eq.~\ref{eq:dNdM} the CMF includes
two other components, the mass ratio $M_0/M$ and the derivative $\dd
S/\dd M$. As a result, transformations that will yield the
appropriate CMF cannot be done using the information imprinted in
$f$ alone, and the functional dependence of $S$ on $M$ should be
important.

We start by searching for the most accurate transformation in $\dW$
which can compensate for the change in cosmology $C$, or descendant
halo mass $M_0$. This means that we only change $\dW$ into
$\widetilde{\dW}$ while keeping the masses of the descendant and
progenitor haloes fixed. In the general case, such a simple
transformation might not allow us to accurately transform the trees.
In practice, it might be a good starting point for many relevant
cases. Our methodology to find time-transformation is as follows:
for a given reference case we use eq. \ref{eq:cmf_fit} to compute
the reference CMF. We then compute the CMF for the target case using
various time-steps $\widetilde{\dW}_i$. The value of
$\widetilde{\dW}$ is chosen from the set of $\widetilde{\dW}_i$ in
order to give the best match between the reference and the target
CMF's.

\begin{figure}
\centerline{ \hbox{ \epsfig{file=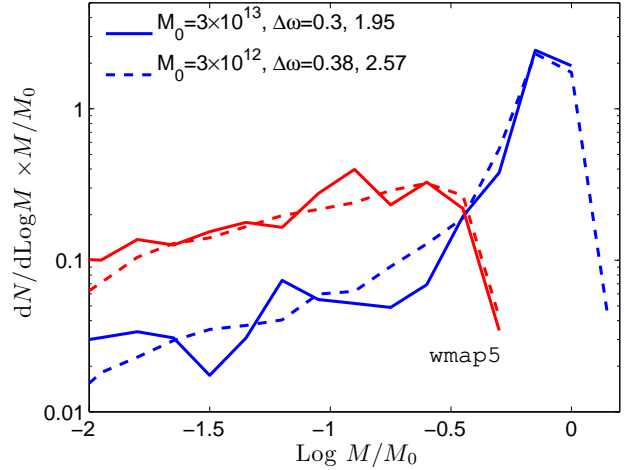,width=9cm} }}
\caption{The conditional mass function (CMF) for different descendant halo masses and at
at different time-steps as indicated.
The figure demonstrates that the transformation $\widetilde{\dW} = 1.3\times\dW$
properly compensates for the change in descendant mass $M_0$.}
  \label{fig:mass_scaling}
\end{figure}

In fig.~\ref{fig:cmf_scaling} we plot the CMF for two sets of
cosmologies where time transformation is being used. This
transformation is relatively accurate, although it can introduce a
large transformation in time (almost a factor of 2 in $\dW$ for one
of the cases tested here). The original differences between the
simulations can be seen in fig.~\ref{fig:cmf_time1} for reference.
Nonetheless, it seems that the transformation in $\dW$ do not
provide a uniform accuracy for all descendant masses. In
fig.~\ref{fig:mass_scaling} we show how changing $\dW$ can
compensate for different descendant halo masses. This indicates that
enhancement of mass resolution can be easily obtained.

Time transformations are limited in accuracy, and can lower the
time resolution of a merger-tree significantly. In addition,
such transformations might stretch the non-markov correlations between consecutive time-steps
in a way that might be different from the behaviour of the simulations.
Our statistical sample does not allow us to explore the last effect in detail.

\begin{figure}
\centerline{ \hbox{ \epsfig{file=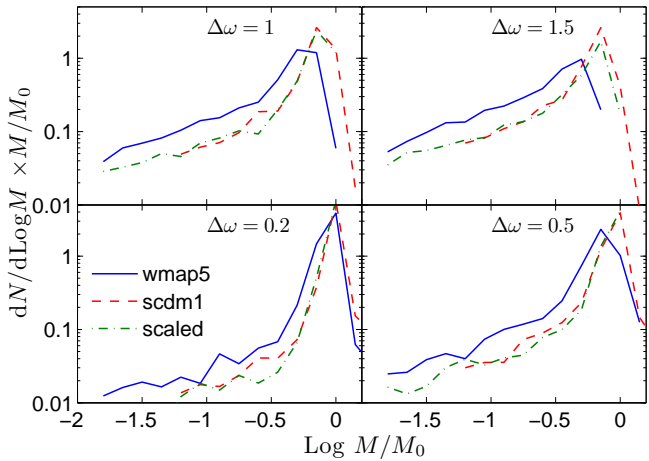,width=9cm} }}
\caption{Scaling the mass in order to compensate for the change in $\dd N/\dd M$.
We use eq.~\ref{eq:mass_scaling} to transform the mass of progenitors from
the wmap5 simulation (solid blue lines) into the scdm1 model (dashed red),
both at the same time $\dW$.
The mass transformation $M_i \rightarrow \widetilde{M}_i$ derives from the global
fit at a \emph{specific} time-step, $\dW=1$.
The same transformation is used to transform the actual progenitor
masses for \emph{all} the time steps used here. The dashed-dotted lines show
the resulting CMF values, which are similar to the target case.
The descendant halo mass is $10^{12}\,\hmsun$.}
  \label{fig:cmf_scaling_mass}
\end{figure}

A less trivial transformation is needed when the time scaling cannot
provide an accurate enough solution. In this case we can transform
the mass of the progenitor haloes in order to yield the same CMF.
For any progenitor mass $M$ we define $\widetilde{M}$ such that the
integral over the number of progenitors will be invariant,
\begin{equation}
\label{eq:mass_scaling}
\widetilde{M}: \;\;\; \int_{\widetilde{M}}^{\widetilde{M}_0} \left[ \frac{\dd N}{\dd M} \right]_t \dd M =
\int_M^{M_0} \left[ \frac{\dd N}{\dd M} \right]_r \dd M \, .
\end{equation}
This equation can be used to find the transformation $M \rightarrow
\widetilde{M}$ numerically using the global fitting function of
eq.~\ref{eq:cmf_fit}. In fig.~\ref{fig:cmf_scaling_mass} we show that
this transformation can yield perfect matching between very
different CMF's. The only limitation in the accuracy is the goodness
of our global fit, which is used to find the mass transformation
above.
As shown in fig.~\ref{fig:cmf_scaling_mass} the \emph{same} mass transformation
is suitable for a large range of time-steps, so the deviations of the CMF fit at small
time-steps do not degrade the accuracy of the mass transformation.

In general, one might use transformation of $\dW$ and mass according to the specific case needed.
Time transformation are much more easy to perform, and should be preferred for cases where the resulting
time resolution is not problematic. Mass transformation can be done as a second step, to increase the
accuracy of the statistics by small variations in mass. For example, as tested above, using the results
of the Millennium simulation \citep{Springel05} with a transformation $\widetilde{\dW}=0.86\dW$ should give
merger trees which are consistent with the cosmological model WMAP5
for a large range of halo masses. For this simple case, the time
transformation is consistent with the naive EPS prediction and scales like $\sigma_8$.

\subsection{Generating Monte-Carlo trees}
\label{sec:mc_trees}

Constructing Monte-Carlo merger trees is doable once an accurate
knowledge of the CMF is given for all time-steps. There are various
algorithms that were applied to the EPS CMF, and can be easily
generalized for other CMF's \citep{Kauffmann93a, Neistein08b,
Zhang08a}. In this sense our fitting function for the CMF might be
very useful for constructing Monte-Carlo trees that fit the results
of \nbss. However, there are some limitations to this approach which
we will emphasize here below.

As stated in the previous section, the quality of our global fit is
poor for small time steps ($\dW<0.5$). On the other hand, the
time-step which is convenient for constructing Monte-Carlo trees is
much smaller, around $\dW=0.1$. Larger time-steps yield many
progenitors in each merger event, and a large uncertainty in the
predicted merging time of the haloes. As a result, we do not have an
accurate fitting function for the CMF, which is appropriate for
generating Monte-Carlo trees. More than that, an accurate fit for
the CMF at small $\dW$ will not provide a solution, because the
behaviour of merger trees is highly non-markov at such small
time-steps. This means that applying the CMF in consecutive
time-steps without proper correlation between steps will generate
large deviations in the merger histories (ND08a).

A possible solution to these problems is to look for a new CMF at
small time-steps that will reproduce the CMF from \nbss\ at big
time-steps. The CMF at small $\dW$ does not have to match \nbss, it
can only be tested by applying it on few consecutive time-steps.
Such a methodology was introduced by ND08a and found to produce good
results. However, the resulting trees are fully markov, differing
from \nbss. As a result of the above complications we think that
this problem deserve more room than what is left here, and we
postpone it to a future work.


\section{Summary and Discussion}
\label{sec:discuss}

We provide an improved description of the merger histories of dark-matter haloes
as measured from \nbss. Using a suite of cosmological simulations, and a range of
descendant halo masses at different redshifts, we provide a robust fitting function
for the conditional mass function of progenitors, the CMF. This fit emphasizes the
self-similar and universal nature of the merger trees and it is more accurate than
earlier attempts \citep{Lacey93,Sheth02,Moreno08,Cole08}.
The improved fit is owing to the functional form used, which generalizes the
EPS formula with a term that breaks the symmetry between the natural time and mass
variables. An improved accuracy is achieved by allowing the best-fit parameters to
explicitly depend on the descendant halo mass.
We note that it might be that other fitting functions will describe the data in a similar
accuracy, and that our formula will deviate from the simulation results at different
halo mass ranges. However, the range in halo mass probed in this work dominates
the contribution to galaxy formation processes.

The accurate fitting function for the CMF can serve as a basis for comparing
different simulations, or different algorithms for construction merger trees.
It can also be used for testing
new modifications to the theory \citep[e.g.][]{Maggiore09} and for constraining the time
evolution of the halo mass function \citep{Tinker08}.
We discuss a specific application of our fitting function, that is to scale
merger trees that are extracted from a given \nbs\ into a different cosmological model,
or different ranges of mass or redshift. We demonstrate that these scalings provide
more accurate merger trees than other methods that utilize Monte-Carlo realizations.
The scaling method conserves the correlation of the progenitor masses between time-steps,
a non-Markovian effect that is hard to mimic in Monte-Carlo generated trees.
It would be worthwhile to study the scaling of substructures along similar lines.
Substructures should be affected mostly by the mass ratio of the subhalo to its host halo,
and by the dynamical time-scale within the host. Once these quantities are conserved,
substructures might be easily scaled as well. Such an analysis of substructures requires
simulations of higher resolution than those used here.

While most predictions of the EPS approximation are inaccurate, we find that the
self-similarity in time proposed by EPS is valid to an accuracy of a few percents.
We show explictly that merger histories are self-similar when using the time variable
$\omega=\delta_c/D(t)$, with $\delta_c\simeq 1.68$ and $D(t)$ the cosmological linear growth rate.
This self-similarity, which has been noticed by \citet{Neistein08a} and \citet{Genel08} in the
Millennium simulation \citep{Springel05}, is verified here for different cosmological models.
We show that this self-similarity implies that $\omega$ can be used for relating merger
trees from different cosmologies.

In this work we used as default a standard halo definition based the \textsc{fof} algorithm with a linking
length corresponding to 20\% of the mean interparticle separation. We tested in comparison
merger trees for haloes defined using a spherical overdensity (SO) method
\citep[for more details on how these trees are defined see][]{Maccio08}, and found
that they are less self-similar in time,
and are thus not as suitable for obtaining a universal fitting function.
Indeed, the \textsc{fof} haloes are known to provide a better match to the theory
\citep{Lacey94}, and they are therefore more commonly used in the relevant literature
\citep[e.g. Sheth \& Tormen 2002\footnote{The merger trees built with the GIF simulation
used by \citet{Sheth02} are described in \citet{Kauffmann99}},][]{Cole08}.

The construction of a merger tree by linking haloes from different simulation snapshots
can be done in different ways, and the results may depend on the method adopted.
A comparison of our results and other results from the literature indicates
that this is not an issue of major concern. For example, \citet{Cole08} does not report
a noticable difference between their CMF results and those of \citet{Bower06}, which use
different algorithms for constructing merger trees.
On the other hand, our progenitor definition is very similar to the one used by \citet{Sheth02},
so the improvement in the fitting quality is not likely to be due to deviations in the
simulation results. Finally, merger trees that were constructed by different methods
may differ at small time steps, but these differences tend to diminish when large time
steps are considered, e.g. the $\Delta z>0.5$ steps used in our analysis
\citep[see e.g.][and their figure 8b]{Genel08}.

Our current analysis of a suite of cosmological simulations confirms
the lognormal nature of the main-progenitor distribution as a function of
the natural mass variable $\sigma^2(M)$, as reported by \citet{Neistein08a} from the Millenium
simulation. The linear dependence of the parameters of the lognormal distribution on the natural time
variable $\log\dW$ is much simpler than the time dependence of the CMF fit described in the
first part of our paper. This motivates a theoretical search for the origin of the lognormal
distribution of main progenitors. Both methods used here to quantify merger-trees as
measured from \nbss\ emphasize the limits of current theories in capturing the details of
dark-matter evolution.


\section*{Acknowledgments}

We thank Raul Angulo, Mike Boylan-Kolchin, Jorge Moreno, Ravi Sheth,
and Simon White for useful discussions. We are grateful to Andreas
Faltenbacher for many helpful comments. Numerical simulations were
performed on the PIA cluster of the Max-Planck-Institut f\"ur
Astronomie and on the PanStarrs2 clusters at the Rechenzentrum in
Garching. EN is supported by the Minerva fellowship. AM was
partially supported by the Astrosim grant 2281. This research was
supported by the German-Israeli-Foundation (GIF).

\bibliographystyle{mn2e}
\bibliography{Eyal_bib}

\appendix

\section{Additional Statistics}
\label{sec:add_stat}


\subsection{Self-similarity in time}
\label{sec:app_time}

\begin{figure}
\centerline{ \hbox{ \epsfig{file=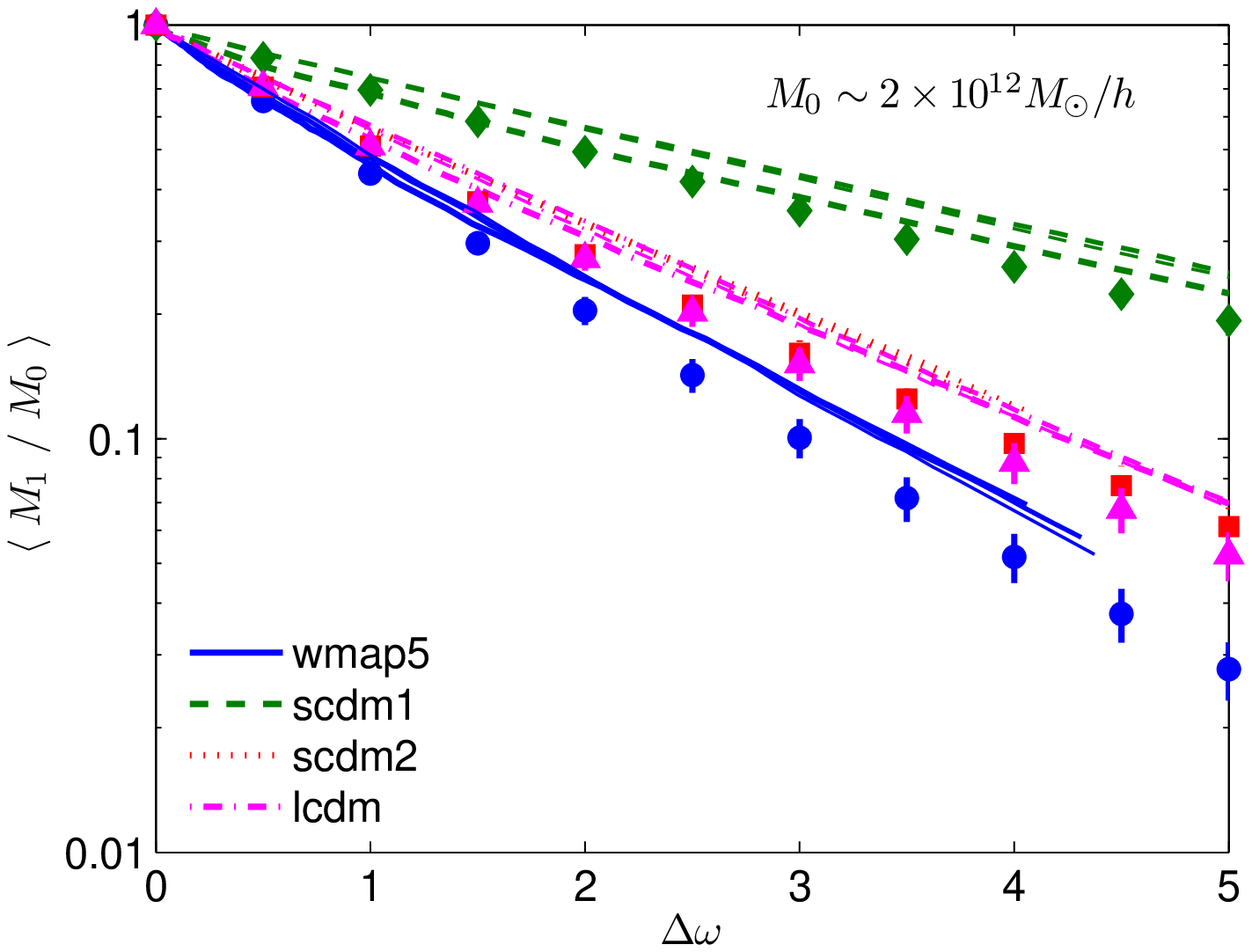,width=9cm} }}
\centerline{ \hbox{ \epsfig{file=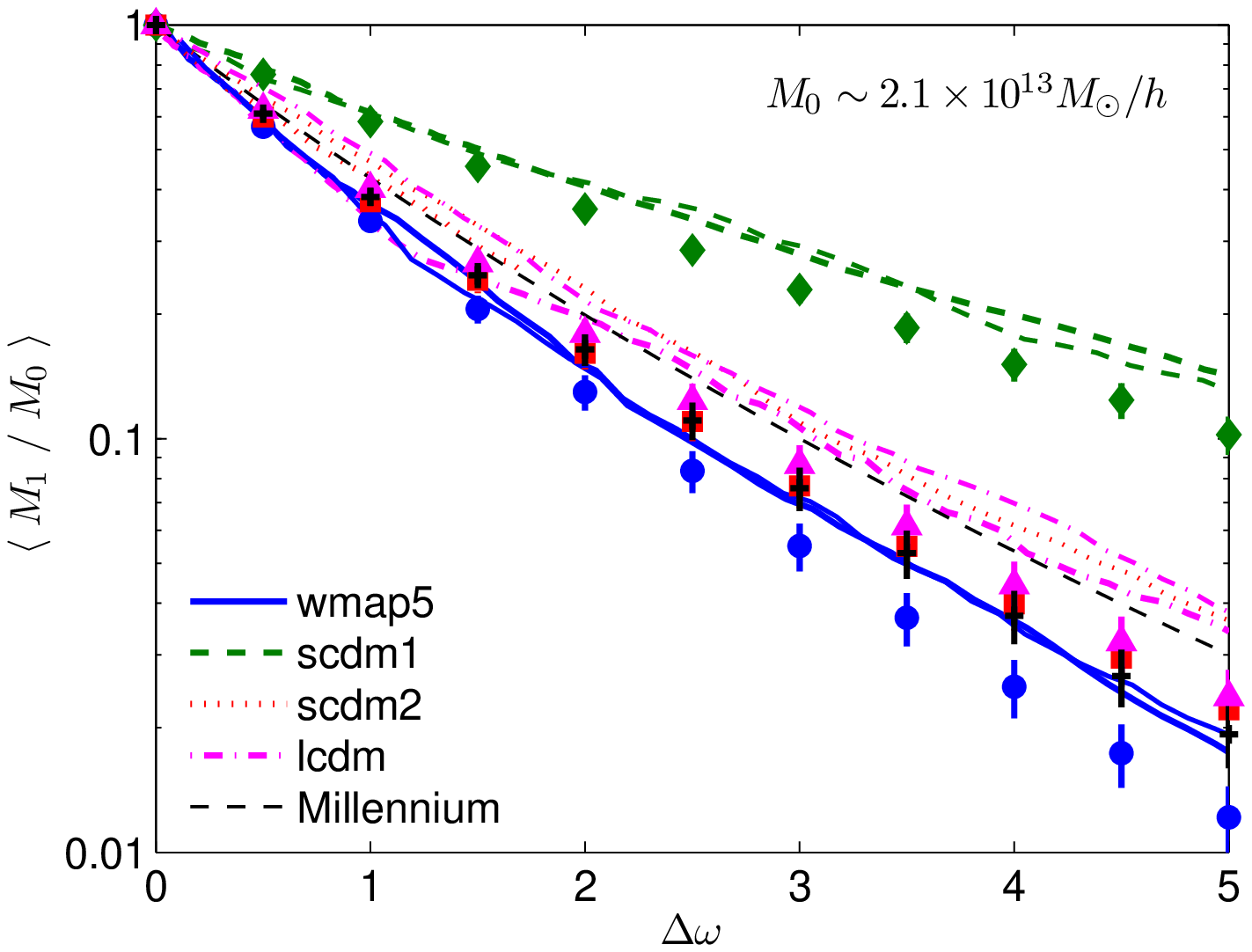,width=9cm} }}
\caption{Average main-progenitor histories at different times. Smooth lines show
main-progenitor histories for descendant halo $M_0$ identified at $\z_0=0,\, 1,\, 2$
($\z_0=2$ was omitted for the high-mass bin due to a small statistical sample). Symbols
and error-bars are the predictions of the EPS formalism following the formula
of \citet{Neistein06}, and are independent on $\z_0$. The symbols shapes are
diamonds, squares, triangles, circles and pluses for the scdm1, scdm2, lcdm, wmap5,
and Millennium simulations respectively. It is evident that self-similarity in
time is valid for all the simulations. The deviations of the EPS formalism with
respect to \nbss\ are similar in all cases, although slight trends
with cosmology and halo mass can be seen.}
  \label{fig:average_m1}
\end{figure}

In this section we provide more details on the accuracy of the
self-similarity in time, as discussed in section \ref{sec:cmf_time}.
In fig.~\ref{fig:average_m1} we plot average main-progenitor
histories for our set of simulations. The histories are plotted for
descendant haloes identified at $\z_0=0,1,2$. Comparing histories
for different $\z_0$ we see small deviations with $\z_0$, reaching
$\sim 20\%$ at $\dW=5$ (for the scdm1 \& lcdm simulations). It
should be kept in mind that cosmic variance is non-negligible in
this plot. We also plot the EPS prediction for the main-progenitor
histories, as given by \citet{Neistein06}. The difference between
the analytical EPS prediction and the results of \nbss\ seem to
change slightly with halo mass. These results are sometimes
different from the study of \citet{Zhao08}. We note that these
authors compared \emph{median} values from simulations against
\emph{average} values from the EPS formalism, as estimated by
\citet{vdBosch02b}. This makes their comparison less accurate than
what is done here.

\begin{figure}
\centerline{ \hbox{ \epsfig{file=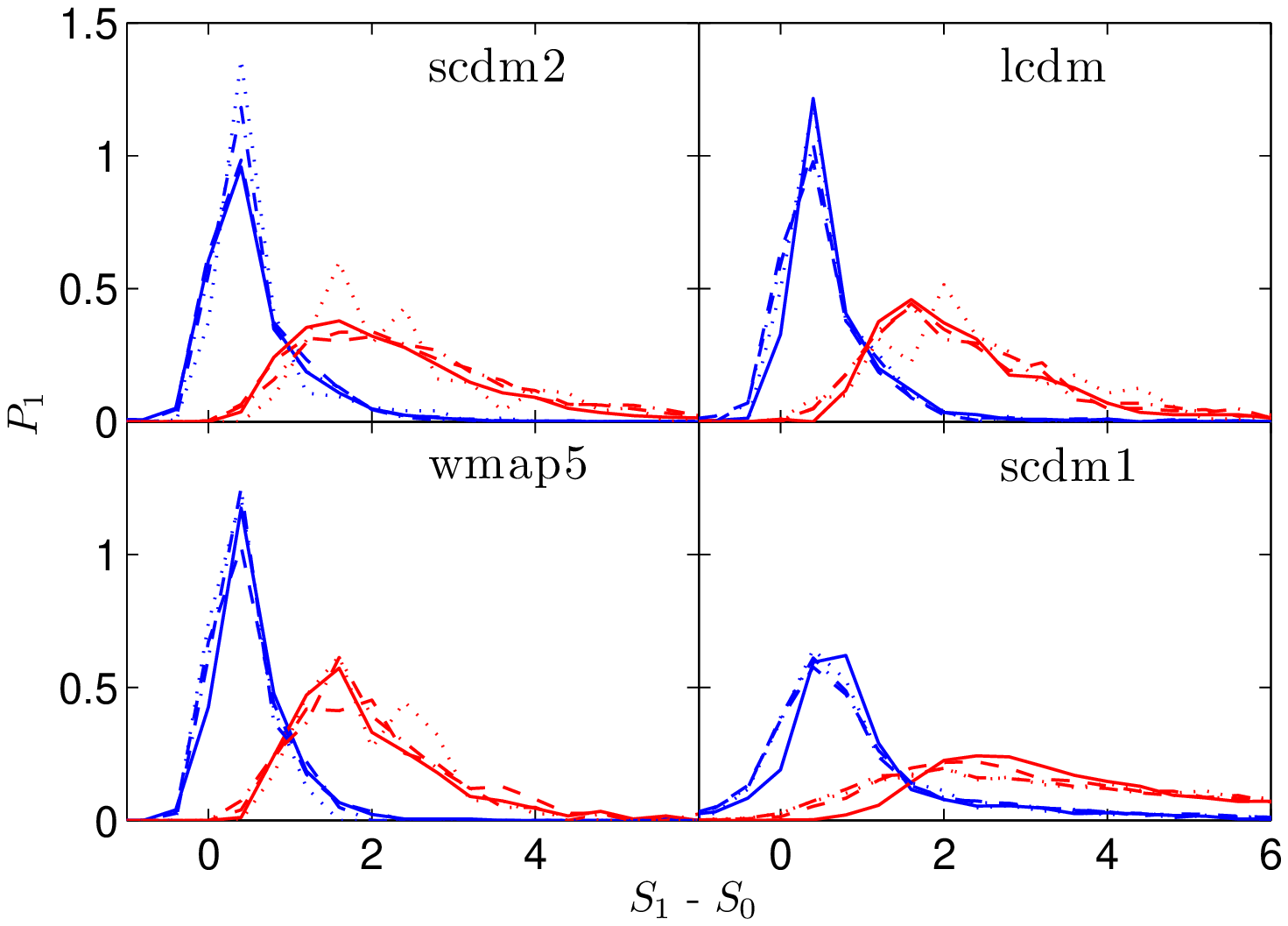,width=9cm} }}
\caption{The full distribution of the main-progenitor mass at different times.
In each panel we plot the results from one simulation as indicated, where the descendant
halo mass $M_0$ is between $10^{12}$ and $10^{13}\,\hmsun$, and it is selected at
$\z_0=0,1,2,3$ (solid, dashed, dashed-dotted, dotted lines respectively). Main-progenitors
are followed backwards in time until $\dW=0.5\,\&\,2$ (blue and red curves respectively).}
\label{fig:p1_time}
\end{figure}

The full distribution of the main-progenitor, $P_1(\dS|S_0,\dW)$, is
plotted in fig.~\ref{fig:p1_time} for descendant haloes selected at
different $\z_0$. Here as well, self-similarity in time is shown to
be accurate with the exception of the scdm1 simulation at $\z_0=0$.

\begin{figure}
\centerline{ \hbox{ \epsfig{file=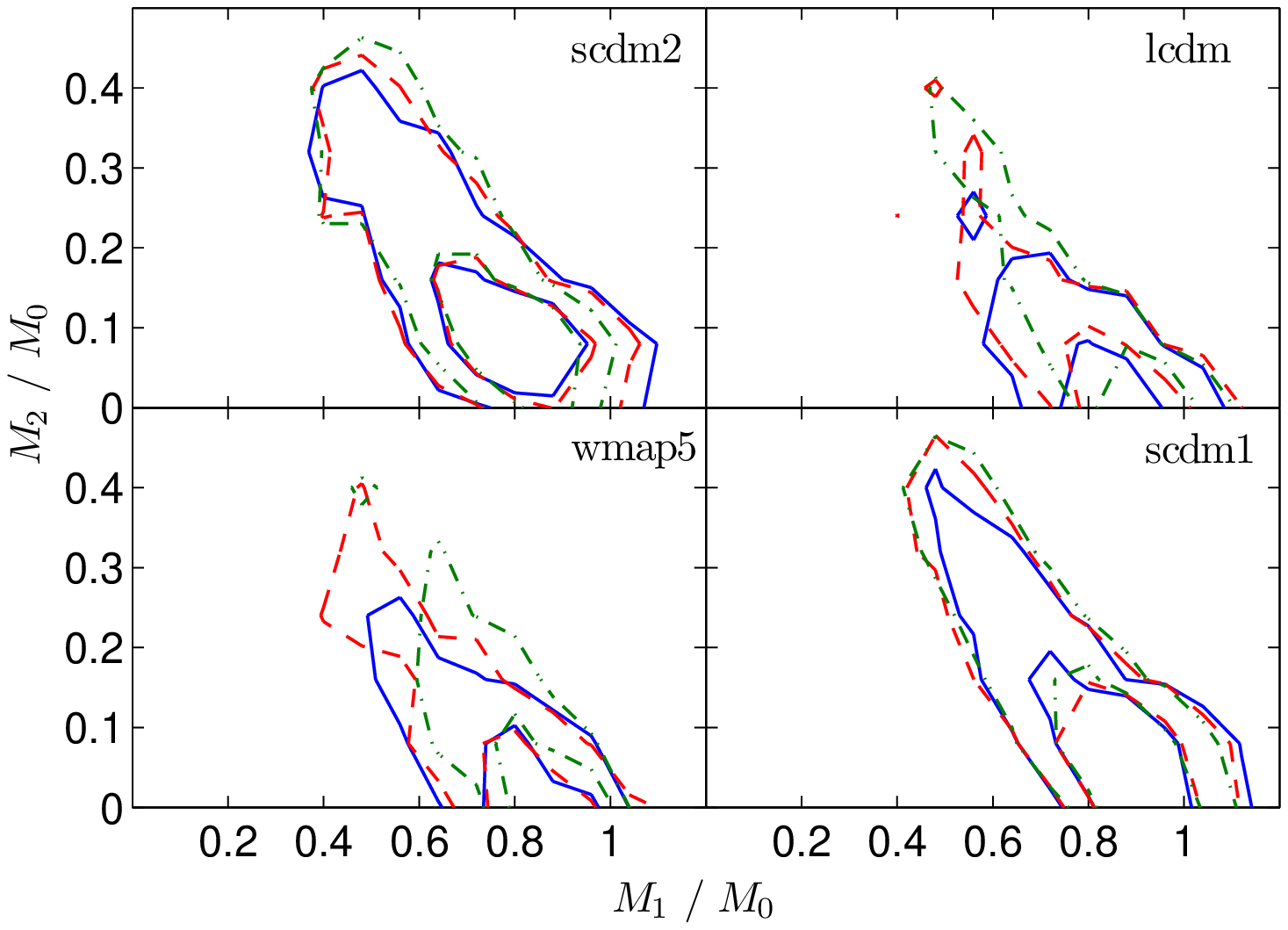,width=9cm} }}
\caption{The mutual distribution of the two most massive progenitors ($M_1,\,M_2$).
Each panel shows results from one simulation as indicated. The descendant halo
$M_0$ is identified at $\z_0=0,1,2$ (plotted as solid, dashed, dashed-dotted lines
respectively). The contour lines are plotted for 7 \& 30\% of the maximum histogram values.
Descendant mass is between $10^{12}$ and $10^{13}$ $\hmsun$, and $\dW=0.3$ in all cases.}
  \label{fig:p12_time_a}
\end{figure}

\begin{figure}
\centerline{ \hbox{ \epsfig{file=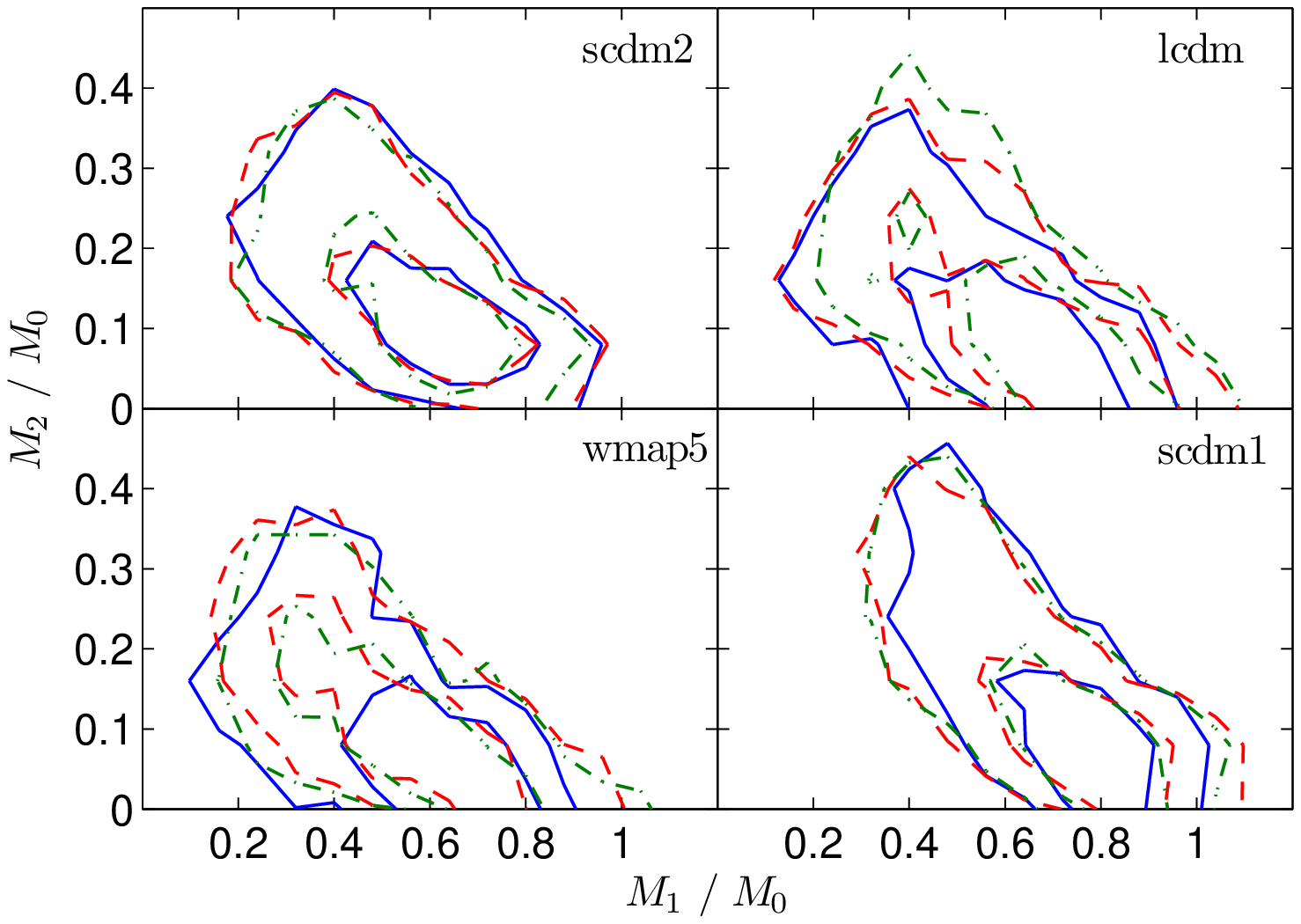,width=9cm} }}
\caption{Same as fig.~\ref{fig:p12_time_a} but with $\dW=0.8$.}
  \label{fig:p12_time_b}
\end{figure}

In figs.~\ref{fig:p12_time_a} \&  \ref{fig:p12_time_b} we plot the
mutual distribution of the two most massive progenitors. We show two
dimensional histograms for descendant haloes selected at
$\z_0=0,1,2$. The results for scdm1 \& scdm2 simulations show
accurate similarity for different $\z_0$. However, the lcdm and
wmap5 show deviations a the level of 10-20\% in mass. These
deviations decrease at larger time steps, so they might be connected
to non-markov effects at small time-steps, and their variation with
redshift. In addition, our treatment of `blacksplash' haloes may
affect the results (see the discussion in section \ref{sec:cmf_time}).

\subsection{Scaling merger trees}
\label{sec:app_scaling}

\begin{figure}
\centerline{ \hbox{ \epsfig{file=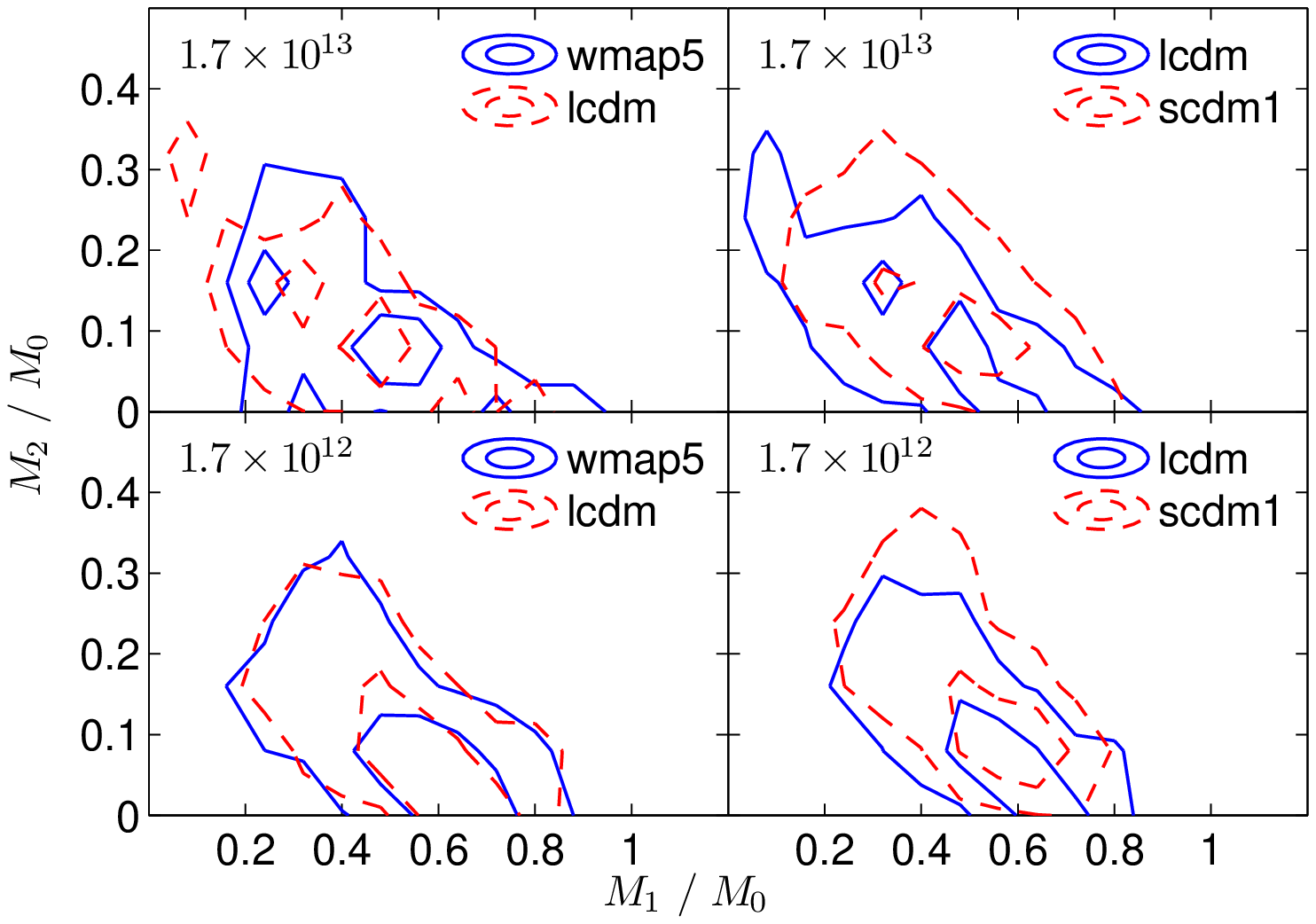,width=9cm} }}
\caption{The mutual distribution of the two most massive progenitors ($M_1,\,M_2$),
for $\dW$ matching. The same $\dW$ transformation is done as in fig.~\ref{fig:cmf_scaling}.
The contour lines are plotted for 13 \& 50\% of the maximum histogram values.
Descendant mass is indicated in units of $\hmsun$, $\dW=1$.}
  \label{fig:p12_rescale}
\end{figure}

\begin{figure}
\centerline{ \hbox{ \epsfig{file=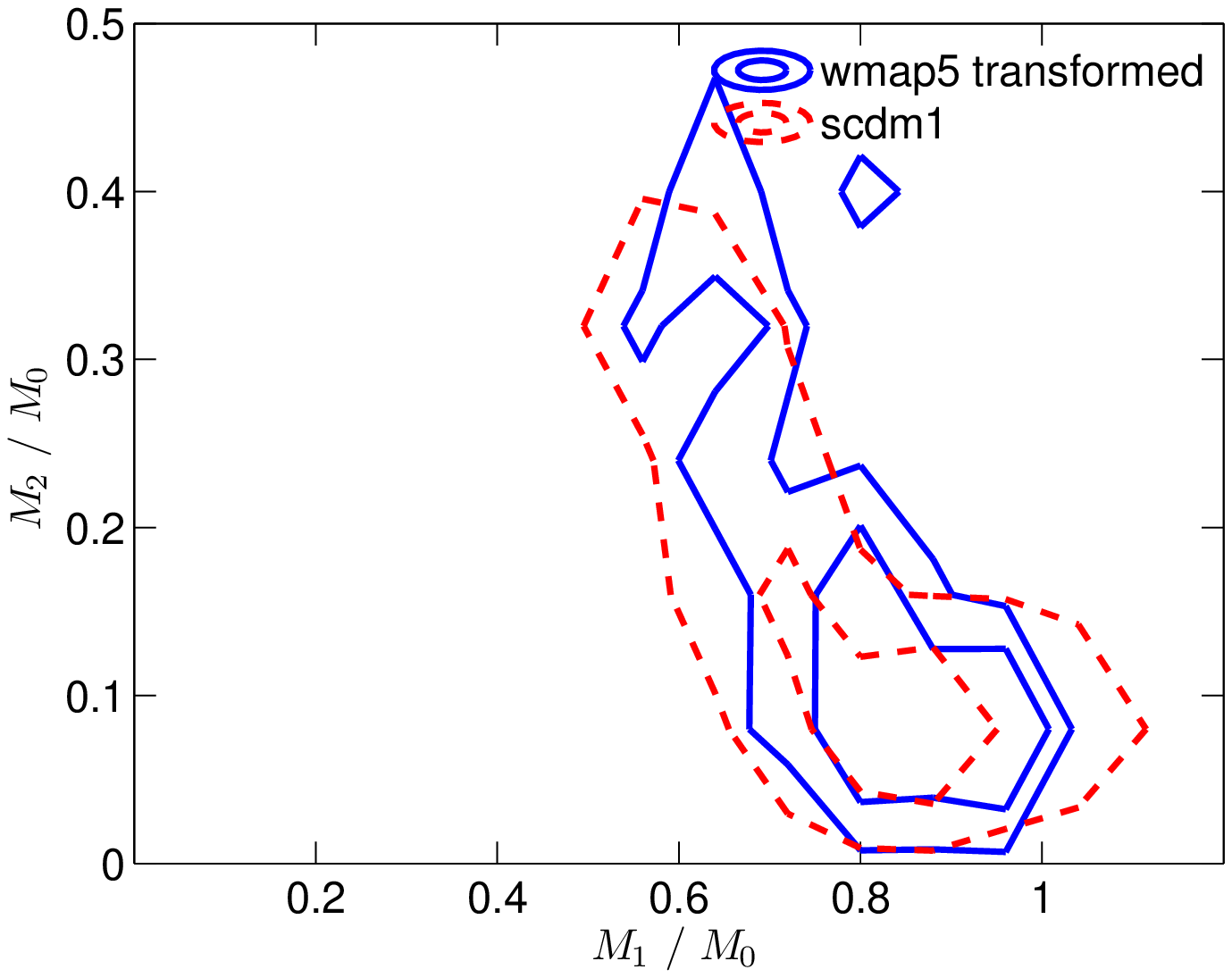,width=9cm} }}
\caption{The mutual distribution of the two most massive progenitors ($M_1,\,M_2$),
for mass matching. Mass transformation for wmap5 was done in the same way as in
fig.~\ref{fig:cmf_scaling_mass}. $\dW=0.3$, $M_0=10^{12}\,\hmsun$. }
  \label{fig:p12_rescalemass}
\end{figure}

In this section we further examine the scaling of merger trees as
it was applied in section \ref{sec:app_scaling}. We only show mutual
distributions of the two most massive progenitors, $(M_1,M_2)$, as
the main-progenitor distribution is highly constrained by the CMF
\citep[see e.g.][]{Neistein06}. In fig.~\ref{fig:p12_rescale} we
examine the time-transformation which was used to generate
fig.~\ref{fig:cmf_scaling}. The results of the mass scaling as
applied in fig.~\ref{fig:cmf_scaling_mass} are shown here in
fig.~\ref{fig:p12_rescalemass}.

\subsection{A global fit to the CMF}
\label{sec:app_cmf}

\begin{figure}
\centerline{ \hbox{ \epsfig{file=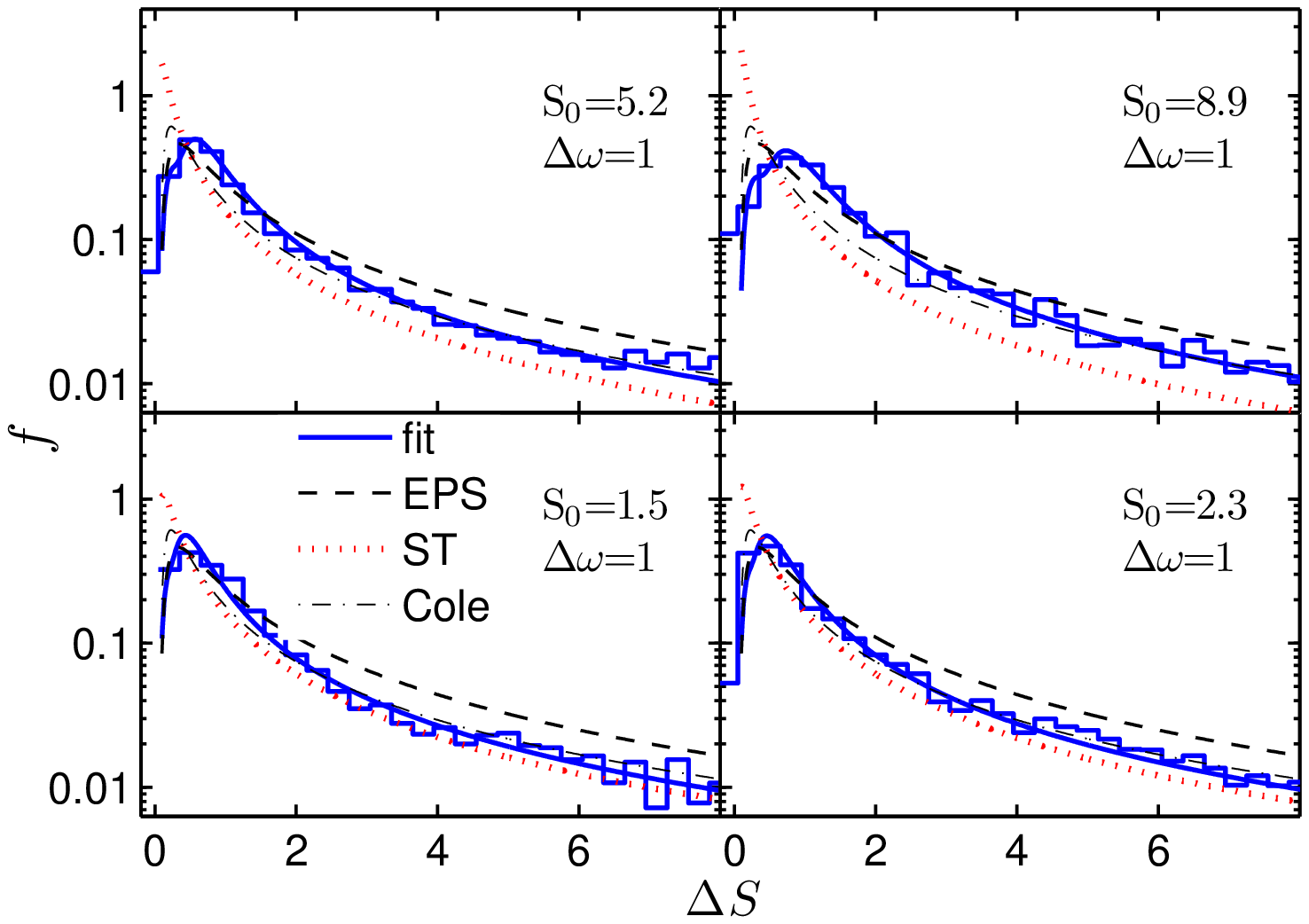,width=9cm} }}
\centerline{ \hbox{ \epsfig{file=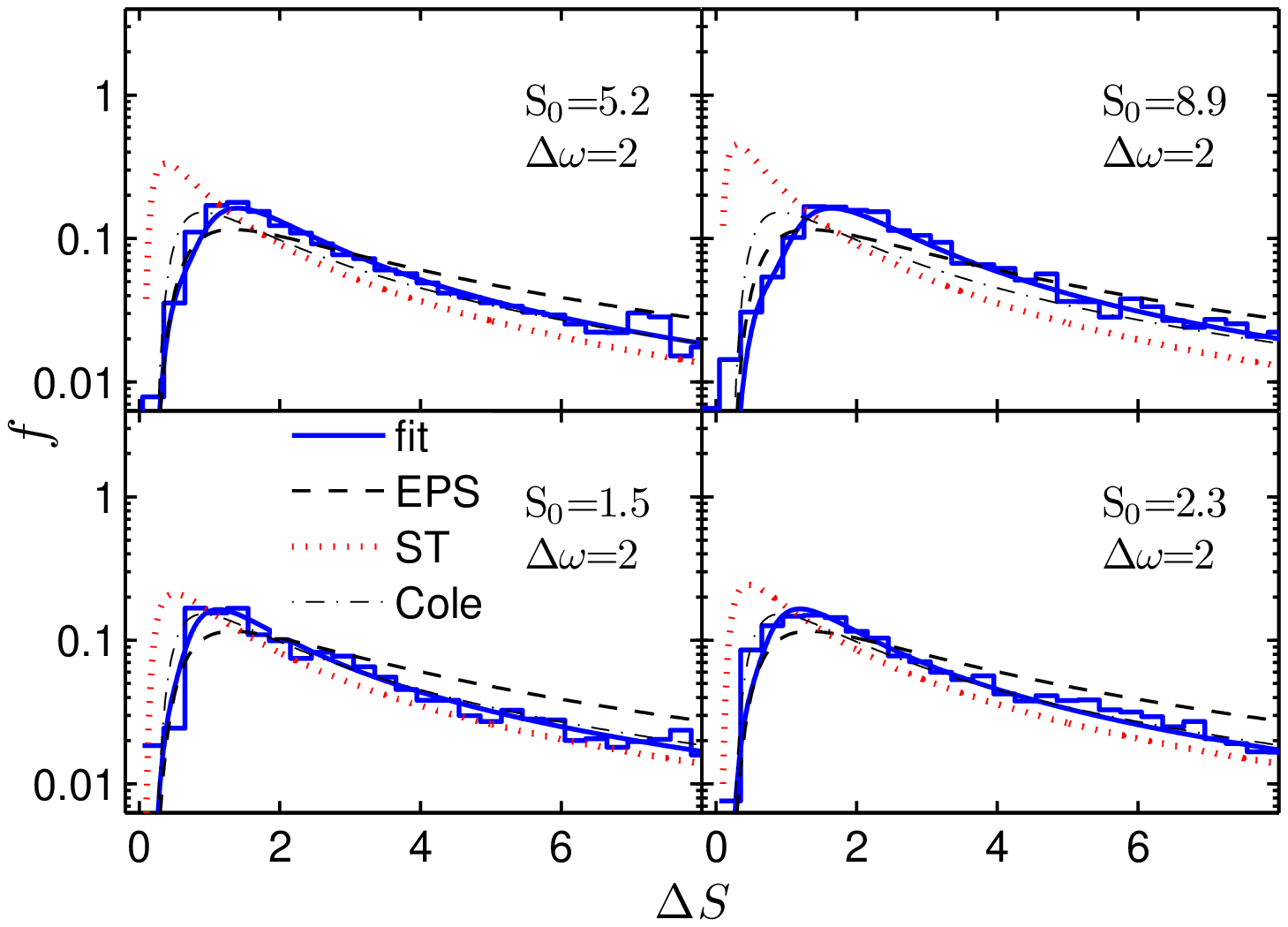,width=9cm} }}
\caption{Similar to fig.~\ref{fig:cmf_fit}, but for different time-steps
as indicated.}
  \label{fig:cmf_fit_app}
\end{figure}

In fig.~\ref{fig:cmf_fit_app} we provide more tests to our global fit,
in a similar way to fig.~\ref{fig:cmf_fit}.

\section{A global fit to the main-progenitor distribution}
\label{sec:app_mp}

\begin{figure}
\centerline{ \hbox{ \epsfig{file=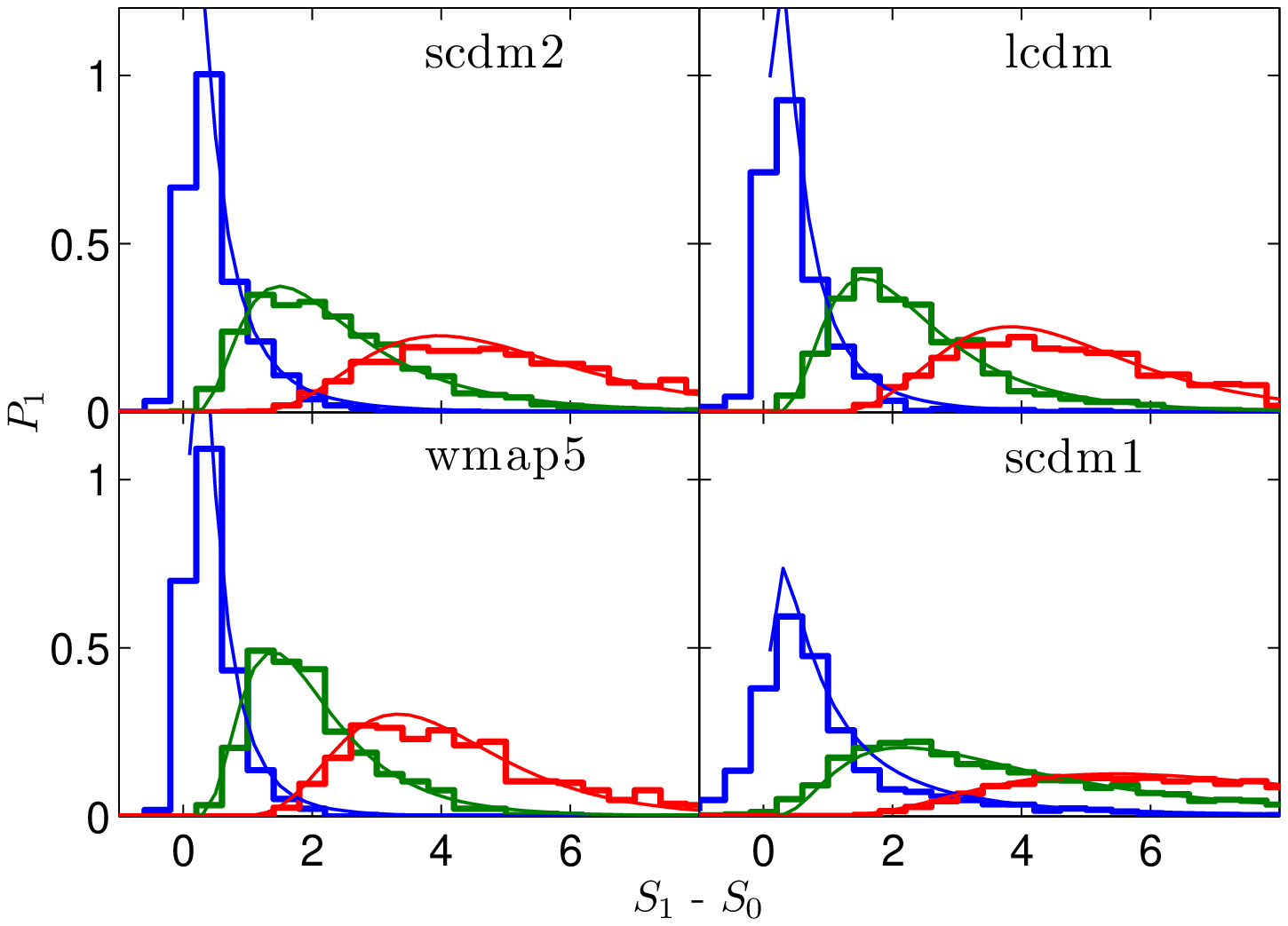,width=9cm} }}
\caption{The full distribution of the main-progenitor for different cosmologies
and time-steps. Descendant haloes are selected at $\z_0=1$ with mass between
$10^{12}$ and $10^{13}\,\hmsun$. Main-progenitor mass is followed backward in
time until $\dW=0.5,\,2,\,4$. Histograms show the simulation data, smooth lines are
generated using our global fit.}
  \label{fig:P1_global}
\end{figure}

In section \ref{sec:mp} we claim that the lognormal distribution can
be easily described for all time-steps and descendant masses using
eqs.~\ref{eq:p1_lognormal} \& \ref{eq:p1_global}. This is shown in
fig.~\ref{fig:P1_global}, where full distributions of the
main-progenitor mass are plotted. In tables \ref{tab:sig_p} \&
\ref{tab:mu_p} we summarize the parameters of the fits.

%
\begin{table}
\caption{The coefficients used for the global lognormal fit,
eqs.~\ref{eq:p1_lognormal} \& \ref{eq:p1_global}. The standard
deviation of $\ln \dS$ obeys the equation $\sigma_p=(a_1\log
S_0+a_2) \log \dW + a_3\log S_0 + a_4$, where $a_i$ are given below
for each cosmology. For the Millennium simulation (MS) we used the
fit given by ND08a, and transformed it to depend on $S_0$.}
\begin{center}
\begin{tabular}{lcccccc}
\hline Simulation & $a_1$ &  $a_2$ & $a_3$ & $a_4$ \\
\hline
wmap5 & -0.333 & -0.321 & 0.0807 & 0.622  \\
scdm1 & -0.0344 & -0.608 & 0.185 & 0.697 \\
scdm2 & 0.760 & -1.085 & 0.184 & 0.668 \\
lcdm  & -1.209 & 0.205 & 0.245 & 0.571 \\
MS  &  0.0135 &  -0.404  &  0.102  &  0.561 \\
\hline
\end{tabular}
\end{center}
\label{tab:sig_p}
\end{table}
\begin{table}
\caption{The coefficients used for the global lognormal fit,
eqs.~\ref{eq:p1_lognormal} \& \ref{eq:p1_global}. The mean of $\ln
\dS$ obeys the equation $\mu_p=(b_1\log S_0+b_2) \log \dW + b_3\log
S_0 + b_4$, where $b_i$ are given below for each cosmology. For the
Millennium simulation (MS) we used the fit given by ND08a, and
transformed it to depend on $S_0$.}
\begin{center}
\begin{tabular}{lcccccc}
\hline Simulation & $b_1$ &  $b_2$ & $b_3$ & $b_4$ \\
\hline
wmap5 & 0.132 & 2.404 & 0.585 & -0.436 \\
scdm1 & -0.8105 & 3.179 & 0.988 & -0.513 \\
scdm2 & 0.418 & 2.366 & 0.999 & -0.647 \\
lcdm  & 0.0788 & 2.418 & 0.671 & -0.434 \\
MS  &  -0.217  & 2.575 & 0.662 & -0.5627 \\
\hline
\end{tabular}
\end{center}
\label{tab:mu_p}
\end{table}

\label{lastpage}

\end{document}